\documentclass[english,tightenlines,10pt,eqsecnum,amsmath,amssymb,aps,prd,superscriptaddress,nofootinbib,showpacs,floats,notitlepage,showkeys]%
              {revtex4-1}
\usepackage{esint}
\usepackage{fmtcount}
\usepackage{threeparttable}
\usepackage{babel}
\usepackage{amstext,amsfonts}
\usepackage{epsfig}
\usepackage{color}
\usepackage{footnote}
\usepackage{amsmath}
\usepackage{fancyhdr}
\usepackage[dvipdfm]{hyperref}
\usepackage{hyperref}

\begin{document}
\title{$f(R,T)$ Cosmological Models in Phase Space}
\author{Hamid Shabani}\email{h\_shabani@sbu.ac.ir}
\author{Mehrdad Farhoudi}\email{m-farhoudi@sbu.ac.ir}
\affiliation{Department of Physics, Shahid Beheshti University,
             G.C., Evin, Tehran, 19839, Iran}
\date{August 18, 2013}
%
\begin{abstract}
\noindent
 We investigate the cosmological solutions of $f(R,T)$
modified theories of gravity for a perfect fluid in a spatially
FLRW metric through the phase space analysis, where $R$ is the
Ricci scalar and $T$ denotes the trace of the energy--momentum
tensor of the matter content. We explore and analyze the three
general theories with the Lagrangians of minimal $g(R)+h(T)$, pure
non--minimal $g(R)h(T)$ and non--minimal $g(R)\left(1+h(T)\right)$
couplings through the dynamical systems approach. We introduce a
few variables and dimensionless parameters to simplify the
equations in more concise forms. The conservation of the
energy--momentum tensor leads to a constraint equation that, in
the minimal gravity, confines the functionality of $h(T)$ to a
particular form, hence, relates the dynamical variables. In this
case, the acceptable cosmological solutions that contain a long
enough matter dominated era followed by a late--time accelerated
expansion are found. To support the theoretical results, we also
obtain the numerical solutions for a few functions of $g(R)$, and
the results of the corresponding models confirm the predictions.
We classify the solutions into six classes which demonstrate more
acceptable solutions and there is more freedom to have the matter
dominated era than in the $f(R)$ gravity. In particular, there is
a new fixed point which can represent the late--time acceleration.
We draw different diagrams of the matter densities (consistent
with the present values), the related scale factors and the
effective equation of state. The corresponding diagrams of the
parameters illustrate that there is a saddle acceleration era
which is a middle era before the final stable acceleration
de~Sitter era for some models. All presented diagrams determine
radiation, matter and late--time acceleration eras very well. The
pure non--minimal theory suffers from the absence of a standard
matter era, though we illustrate that the non--minimal theory can
have acceptable cosmological solutions.
\end{abstract}

\pacs{04.50.Kd; 95.36.+x; 98.80.-k; 98.80.Jk}
 \keywords{Cosmology; $f(R,T)$ Gravity; Dark Energy; Dynamical Systems Approach; Modified Theories of Gravity.}
\maketitle
\section{Introduction}\label{intro}
Since the birth of general relativity (GR) in 1915, the theory has
faced the appearance of new ideas on changing or even replacing it
in favor of an alternative one\footnote{For example, see
Refs.~\cite{fard,fR4} and references therein.} which could solve
different aspects or at least some parts of its incompleteness and
shortcomings. These novel ideas mainly consist of some
modifications or generalizations which would challenge GR in a
geometrical background. Some of these theories introduce extra
dimensions e.g., the Kaluza--Klein theories~\cite{kaluza} and
braneworld scenarios~\cite{brane}. Other alternatives are
scalar--tensor theories e.g., the Brans--Dicke theory~\cite{brans}
and higher--order/modified
gravities~\cite{fR4,fare,NO2007,fR1,fR2,fR3,NO2011,CliftonEtal}.
Another possibility is to introduce some new cosmic fluids e.g.,
dark matter~\cite{dmatt1,dmatt2,dmatt3,dmatt4,dmatt5}, which
should give rise to the clustered structures, and dark
energy~\cite{dener1,dener2,dener3,dener4,BambaEtal}, which is
responsible for the observed accelerated expansion of the
universe. In particular, the following issues can be addressed as
GR insufficiencies. Fundamentally, incompatibility with the
quantum theory and observationally, inability to explain the
flatness of galaxy rotation curves~\cite{rotca1,rotca2}. Also, the
existing problems of the isotropic and homogeneous cosmological
solution of GR (the standard big--bang cosmology) such as the
horizon and the flatness problems~\cite{Cospr1}, and the absence
of solution(s) including the well--accepted states of cosmological
evolution in the past and future. That is, an accelerating phase
solution prior to the radiation--dominated era e.g.,
inflation~\cite{infla1,infla2,infla3,infla4,infla5}, and an
acceleration phase needed to explain the present accelerated
expansion observed by, e.g., the supernova Ia
observations~\cite{supno1,supno2,supno3,supno4,supno5}, the
large--scale structure (LSS)~\cite{Lss1,Lss2}, the baryon acoustic
oscillations (BAO)~\cite{BAO1,BAO2,BAO3}, the cosmic microwave
background radiation (CMBR)~\cite{CMBR1,CMBR2,CMBR3} and the weak
lensing~\cite{wlen1}. Of course, in spite of the above
deficiencies, matching the experimental results for the precession
of the Mercury orbit~\cite{damour,ctest1,ctest2}, the
Lense--Thirring gravitomagnetic precession~\cite{ls-th1}, the
gravitational deflection of light by the
sun~\cite{damour,ctest1,ctest2} and the gravitational
redshift\footnote{However, any relativistic theory of gravitation
consistent with the principle of equivalence will predict a
redshift.}~\cite{damour,ctest1,ctest2} are the excellent successes
of GR. On the whole, the results of Einstein's theory when
considering the development of a general phenomenological
framework, i.e. the PPN--formalism, determine that it is the best
known metric theory of gravity~\cite{ctest1}.

Among the extended theories of gravity, there are at least two
main motivations\footnote{See, e.g., Ref.~\cite{fare}.}\
 for employing the higher--order
gravities, i.e., those in which the Einstein--Hilbert action is
modified by higher--order curvature invariants with respect to the
Ricci scalar. The first motivation has a theoretical background
and is related to the non--renormalizability of
GR~\cite{nonno1,nonno2} and to the fact that GR cannot be
quantized conventionally. Regarding this  issue, some authors have
shown that the inclusion of higher--order terms can solve this
problem~\cite{norma1,norma2}. The other motivation is related to
the recently achieved data in astrophysics and cosmology. Two
contemporary evidences, which are referred to as dark matter and
dark energy, have challenged our knowledge about the universe and
have accounted for the first signals of GR breakdown. It is also
worth mentioning that the concordance or $\Lambda$CDM
model~\cite{LCDM}, the simplest model which adequately fits the
present observations, supported by an inflation
scenario\rlap,\footnote{Such a scenario is needed because the
accelerated phase in the very early universe should end to connect
to a radiation dominated phase; however, a cosmological constant
cannot fulfill this requirement~\cite{Cospr1}.}\ can eventuate an
accelerating phase in the very early and the late universe.
However, the $\Lambda$CDM model suffers from the well--known
cosmological problem originating in pertaining the cosmological
constant to the vacuum energy~\cite{Cos.pro1,Cos.pro2,Cos.pro3}.
That is, the cosmological constant is tremendously small with
respect to the vacuum energy that is defined in particle physics.
With regard to this, a mechanism is needed to get such a small
value to match the present observations, e.g., dynamical dark
energy models contain such mechanisms~\cite{Dyn.DE1,Dyn.DE2}.

$f(R)$ gravities, as the simplest family of the higher--order
gravities, are obtained by replacing the Ricci scalar with a
function $f(R)$ in the Einstein--Hilbert action. Generally, every
new gravity theory, when is introduced as an alternative to GR,
would be tested in two realms. That is, the weak field tests, i.e.
those that can elaborate whether the theory leads to the known
solar system observations, and the cosmological tests which
inspect the theory aiming to find at least a solution that matches
the present accelerated expansion observations. $f(R)$ gravity
theories are~not excepted from these examinations. In these
issues, a few authors have claimed that the solar system tests
rule out most $f(R)$ models~\cite{Rulout1,Rulout2,Rulout3}, though
others do~not agree with these results~\cite{fR1}. However, these
issues do~not seem to be settled completely; see, e.g.,
Refs.~\cite{fR2,CliftonEtal}. Despite the results of local gravity
tests, one can still look at these theories for cosmological
solutions as an independent criterion~\cite{indepe1}. In this
sense, these theories can be considered as $f(R)$ dark energy
models implying that they can play the role of dark energy without
using a cosmological constant, i.e., they can encompass these
problems in a self--consistent scheme. Nevertheless, in addition
to $f(R)$ gravities, there are numerous alternative gravity
theories that claim to cure the problems of dark energy and
inflation, in which up to now, the most physical contents of these
theories have been widely explored; see, e.g.,
Refs.~\cite{fR4,BambaEtal}.

In this work, we purpose to study the cosmology of the so--called
$f(R,T)$ gravity, first introduced in Ref.~\cite{harko} and then,
studied in
Refs.~\cite{frtco,frtec,frttd1,frttd2,frttd3,frtes,JamilEtal,frtsp}.
The theory of $f(R,T)$ gravity generalizes $f(R)$ theories of
gravity by incorporation of the trace of energy--momentum tensor
in addition to the Ricci scalar. The justification for the
dependence on $T$ comes from inductions arising from some exotic
fluid and/or quantum effects (conformal anomaly\footnote{See,
e.g., Refs.~\cite{nonno1,farc}.}). Actually, this induction point
of view adopts or links with the known proposals such as
geometrical curvature inducing matter, a geometrical description
of physical forces, and a geometrical origin for the matter
content of the universe\rlap.\footnote{See, e.g.,
Refs.~\cite{fard,farc} and references therein.}\
 In Ref.~\cite{harko}, the field equations of
some particular models are presented, and specially, scalar field
models $f(R,T^{\phi})$ are analyzed in detail with a brief
consideration of their cosmological implications. Also, the
equation of motion of the test particle and the Newtonian limit of
this equation are further analyzed in Ref.~\cite{harko}. Up to
now, the issues which have been investigated along with this
modified theory are the energy conditions~\cite{frtec}, the
thermodynamics~\cite{frttd1,frttd2,frttd3}, anisotropic
cosmology~\cite{frtes}, the cosmology in which the representation
employs an auxiliary scalar field~\cite{frtco}, the reconstruction
of some cosmological models~\cite{JamilEtal} and the scalar
perturbations~\cite{frtsp}. Also, a further generalization of this
theory has been proposed recently in
Refs.~\cite{OdintsovSaez,HaghaniEtal}. Incidentally, in the
literature, authors have worked on a theory of gravity titled
``the $f(T)$ gravity'' (see, e.g., Ref.~\cite{IorioEtal} and
references therein), where $T$ in this theory is the torsion
scalar arising from the torsion tensor in a similar way as the
curvature scalar arises from the curvature tensor, and which is
completely different from $f(R,T)$ gravity.

On the other hand, $f(R,T)$ gravity may be considered as a
correction or generalization of $f(R)$ gravities as long as the
cosmological considerations are concerned. In this regard, we
employ $f(R,T)$ gravity in an extended version and almost similar
to the terminology employed for the $f(R)$ gravity in
Ref.~\cite{amend}. Hence, in the forthcoming sections, we
theoretically investigate the cosmological solutions of $f(R,T)$
models and compare the results with the corresponding $f(R)$
cosmological models of Ref.~\cite{amend}. To support the
theoretical results, we also obtain the numerical solutions. In
Sec.~\ref{Fieldequation}, we derive the equations of motion (EOM)
for $f(R,T)$ gravity and show that in this case, the conservation
of the energy--momentum tensor leads to a constraint equation that
must be satisfied by any function of $f(R,T)$. For example, for a
minimal coupling form $f(R,T)=g(R)+h(T)$, this constraint
restricts the form of function $h(T)$. Then, we introduce a number
of variables to simplify the equations in more concise forms for
the later applications. In Sec.~\ref{minimalsection}, we analyze
the cosmological solution through the dynamical systems approach.
In this section, we consider the minimal combination, and obtain
the corresponding solutions and the conditions for the existence
of acceptable solutions. In Sec.~\ref{case studies}, we
investigate the numerical results for several functions of
$f(R,T)$ in order to support the theoretical outcomes. In
Secs.~\ref{non1} and~\ref{non2}, we extend the discussion to the
non--minimal combinations and finally, we summarize the obtained
results in the last section.
\section{Field equations of the theory}\label{Fieldequation}
In this section, we obtain the field equations of $f(R,T)$ gravity
and then, introduce some dimensionless variables to simplify the
corresponding equations. The action can be written in the form
\begin{align}\label{action}
S=\int \sqrt{-g} d^{4} x \left[\frac{1}{16 \pi G} f(R,T^{\textrm{(m)}})+L^{\textrm{(m)}}+L^{\textrm{(rad)}} \right],
\end{align}
where $R$ is the Ricci scalar, $T^{\textrm{(m)}}\equiv g^{\mu \nu}
T^{\textrm{(m)}}_{\mu \nu}$ is the trace of the energy--momentum
tensor, the superscript $m$ stands for the dust matter,
$f(R,T^{\textrm{(m)}})$ is an arbitrary function of the Ricci
scalar and $T^{\textrm{(m)}}$, $L^{\textrm{(m)}}$ and
$L^{\textrm{(rad)}}$ are the Lagrangians of the dust matter and
radiation, $g$ is the determinant of the metric and we set $c=1$.
As $T^\textrm{(rad)}=0$, the trace of the radiation
energy--momentum tensor does~not play any role in the function of
$f(R, T^{\textrm{(m)}})$ and henceforth, from now on we drop the
superscript $m$ from the trace $T^{\textrm{(m)}}$ unless it is
necessary. The energy--momentum tensor is usually defined as the
Euler--Lagrange expression of the matter Lagrangian, i.e.,
\begin{align}\label{Energy}
T_{\mu \nu}\equiv-\frac{2}{\sqrt{-g}} \frac{\delta\left[\sqrt{-g} (L^{\textrm{(m)}}+L^{\textrm{(rad)}})\right]}{\delta g^{\mu \nu}},
\end{align}
and if one assumes that both the Lagrangians depend only on the
metric and not on its derivatives, one will get
\begin{align}
T_{\mu \nu}=g_{\mu \nu} [L^{\textrm{(m)}}+L^{\textrm{(rad)}}]-2\frac{\partial [L^{\textrm{(m)}}+L^{\textrm{(rad)}}]}{\partial g^{\mu \nu}}.
\end{align}
By the metric variation of action (\ref{action}), the field
equations are\footnote{By the variational (functional) derivative
procedure (see, e.g., Refs.~\cite{fard,loru}) and employing the
Palatini equation (identity), one can usually derive field
equations; nevertheless, one can consult the detailed derivation
of these field equations in Ref.~\cite{harko}.}
\begin{align}\label{field tensor}
&F(R,T) R_{\mu \nu}-\frac{1}{2} f(R,T) g_{\mu \nu}+\Big{(} g_{\mu \nu}
\square -\triangledown_{\mu} \triangledown_{\nu}\Big{)}F(R,T)=\Big{(}8
\pi G+ {\mathcal F}(R,T)\Big{)} T^{\textrm{(m)}}_{\mu \nu}+8 \pi G  T^\textrm{(rad)}_{\mu \nu},
\end{align}
where it is helpful to define the derivatives with respect to the trace $T$ and the Ricci scalar $R$ as
\begin{align}
{\mathcal F}(R,T) \equiv \frac{\partial f(R,T)}{\partial T}~~~~~~~~~~\mbox{and}~~~~~~~~~~F(R,T) \equiv \frac{\partial f(R,T)}{\partial R},
\end{align}
and we have used
\begin{align}
g^{\alpha \beta} \frac{\delta T^{\textrm{(m)}}_{\alpha \beta}}{\delta g^{\mu \nu}}=-2T^{\textrm{(m)}}_{\mu \nu}.
\end{align}
Also, by contracting equation (\ref{field tensor}), we have
\begin{align}\label{trace tensor}
F(R,T)R+3 \square F(R,T)-2f(R,T)=\Big{(}8 \pi G+ {\mathcal F}(R,T)\Big{)}T.
\end{align}

Now, in this model, we assume a perfect fluid and a spatially flat
Friedmann--Lema\^{\i}tre--Robertson--Walker (FLRW) metric
\begin{align}\label{metricFRW}
ds^{2}=-dt^{2}+a^{2}(t) \Big{(}dx^{2}+dy^{2}+dz^{2}\Big{)},
\end{align}
where $a(t)$ is the scale factor. Let us rewrite
equation~(\ref{field tensor}) in a standard form similar to GR,
i.e.,
\begin{align}\label{FRW}
G_{\mu \nu}=\frac{8 \pi G}{F(R,T)} \left(T^{\textrm{(m)}} _{\mu \nu}+T^{\textrm{(rad)}} _{\mu \nu}+T^{\textrm{(eff)}} _{\mu \nu}\right),
\end{align}
where
\begin{align}\label{Generalized einstein}
T^{\textrm{(eff)}} _{\mu \nu}\equiv \frac{1}{8 \pi G}\left[\frac{1}{2}
\Big{(}f(R,T)-F(R,T)R\Big{)}g_{\mu \nu}+\Big{(} \triangledown_{\mu}
\triangledown_{\nu}-g_{\mu \nu} \square\Big{)}F(R,T)+{\mathcal F}(R,T) T^{\textrm{(m)}}_{\mu \nu}\right].
\end{align}
Regarding the Bianchi identity\rlap,\footnote{It is well--known
that the use of the action principle and the principle of general
invariance allows immediate connections between symmetry
principles and conservation laws to be established as inner
identities. That is, the metric variation of each Lagrangian
density (as a scalar density) of weight one, which is a function
of the metric and its derivatives, makes the covariant divergence
of the Euler--Lagrange expression of the Lagrangian density
identically vanish, e.g., $\nabla^{\mu}T_{\mu \nu}\equiv 0$; see
any text on gravitation, e.g., Ref.~\cite{deInverno}.}\
 obviously in
$f(R,T)$ gravity, the above effective energy--momentum tensor
is~not conserved. Thus, by applying the conservation of the
energy--momentum tensor of all matter and knowing that
$\nabla^{\mu}T^{\textrm{(m)}}_{\mu
\nu}=0=\nabla^{\mu}T^{\textrm{(rad)}}_{\mu \nu}$, the following
constraint must hold. That is
\begin{align}\label{source}
\frac{3}{2} H(t) \mathcal F(R,T)=\dot {\mathcal F}(R,T),
\end{align}
where dot denotes the derivative with respect to the cosmic time $t$ and
$H(t)=\dot{a}(t)/a(t)$ is the Hubble parameter. Obviously this relation
leads to some restrictions on the functionality of $f(R,T)$, as we
shall see in the next section. Equations (\ref{field tensor}) and
(\ref{trace tensor}), by assuming metric (\ref{metricFRW}), give
\begin{align}\label{first}
3H^{2}F(R,T)+\frac{1}{2} \Big{(}f(R,T)-F(R,T)R\Big{)}+3\dot{F}
(R,T)H=\Big{(}8 \pi G +{\mathcal F}
(R,T)\Big{)}\rho^{\textrm{(m)}}+8 \pi G\rho^{\textrm{(rad)}}
\end{align}
as the Friedmann--like equation, and
\begin{align}\label{second}
2F(R,T) \dot{H}+\ddot{F} (R,T)-\dot{F} (R,T) H=-\Big{(}8 \pi G
+{\mathcal F} (R,T)\Big{)}\rho^{\textrm{(m)}}-\frac{32}{3} \pi
G\rho^{\textrm{(rad)}}
\end{align}
as the Raychaudhuri--like equation.

In the following, we assume those functions of $f(R,T)$ that can
be explicitly written as combinations of a function $g(R)$ and a
function $h(T)$, e.g., $f(R,T)=g(R)h(T)$; however, due to the
constraint equation (\ref{source}), their forms would be
restricted. Now, it is convenient to introduce a few dimensionless
independent variables to simplify the obtained equations in the
phase space, used in the following sections. These variables are
defined as
\begin{align}
&x_{1}\equiv-\frac{\dot{g'}(R)}{H g'(R)},\label{varx1}\\
&x_{2}\equiv-\frac{g(R)}{6 H^{2} g'(R)},\label{varx2}\\
&x_{3}\equiv \frac{R}{6 H^{2}}=\frac{\dot{H}}{H^{2}}+2,\label{varx3}\\
&x_{4}\equiv-\frac{h(T)}{3 H^{2} g'(R)},\label{varx4}\\
&x_{5}\equiv\frac{8\pi G \rho^{\textrm{(rad)}}}{3 H^{2} g'(R)},\label{varx5}\\
&x_{6}\equiv-\frac{Th'(T)}{3 H^{2} g'(R)}\label{varx6},
\end{align}
where the prime denotes the ordinary derivative with respect to
the argument and we have used $R=6(\dot{H}+2H^2)$ for metric
(\ref{metricFRW}). However, it will be shown in
Sec.~\ref{minimalsection} that these six variables of the phase
space reduce to five independent variables once the constraint
equation (\ref{source}) is applied. One may also define some other
dimensionless parameters that can play the role of parametrization
in the determination of the function $f(R,T)$, namely,
\begin{align}
&m \equiv \frac{R g''(R)}{g'(R)},\label{parameterm}\\
&r \equiv -\frac{R g'(R)}{g(R)}=\frac{x_{3}}{x_{2}}\label{parameterr},\\
&n \equiv \frac{T h''(T)}{h'(T)}\label{n},\\
&s \equiv \frac{T
h'(T)}{h(T)}=\frac{x_{6}}{x_{4}},\label{parameters}
\end{align}
where $g(R)\neq \mbox{\textrm{constant}}$ and $h(T)\neq
\mbox{\textrm{constant}}$. Note that, generally, we
have\footnote{Actually, in principle, one can derive $R$ and $T$
from definitions (\ref{parameterr}) and (\ref{parameters}) in
terms of $r$ and $s$, respectively. Hence, one gets $m=m(r)$ and
$n=n(s)$.} $m=m(r)$ and $n=n(s)$.

One knows that from the Friedmann equations in GR with the FLRW
metric, the relation $w= p/\rho=-1-2\dot{H}/3H^{2}$ for the
equation of state is obtained. Analogously, if one correspondingly
defines an effective equation of state (for an effective pressure
and an effective energy density) as $w^{\textrm{(eff)}}=
p^{\textrm{(eff)}}/ \rho^{\textrm{(eff)}}
\equiv-1-2\dot{H}/3H^{2}$ then, one will obtain the effective
equation of state as follows. At first, let us redefine equations
(\ref{first}) and (\ref{second}) in a more useful manner for
matching with the SN Ia observations, as
\begin{align}\label{redefinition1}
3A H^2=8\pi G (\rho^{\textrm{(m)}}+\rho^{\textrm{(rad)}}+\rho^{\textrm{(DE)}})
\end{align}
and
\begin{align}\label{redefinition2}
-2A\dot{H}=8\pi G \Big{(}\rho^{\textrm{(m)}}+(4/3)\rho^{\textrm{(rad)}}+\rho^{\textrm{(DE)}}+p^{\textrm{(DE)}}\Big{)},
\end{align}
where $A$ is a constant and $\rho^{\textrm{(DE)}}$ and
$p^{\textrm{(DE)}}$ denote the density and the pressure of the
dark energy, defined as
\begin{align}\label{DEd}
8\pi G\rho^{\textrm{(DE)}}\equiv{\mathcal F}\rho^{\textrm{(m)}}-3\dot{F}(R,T)H-\frac{1}{2}\Big{(}f(R,T)-F(R,T)R\Big{)}+3H^2(A-F)
\end{align}
and
\begin{align}\label{DEp}
8\pi Gp^{\textrm{(DE)}}\equiv\ddot{F}(R,T)+2\dot{F}(R,T)H+\frac{1}{2}\Big{(}f(R,T)-F(R,T)R\Big{)}-(2\dot{H}+3H^2)(A-F).
\end{align}
Thus, the equation of state parameter for the dark energy is given as $w^{\textrm{(DE)}}\equiv p^{\textrm{(DE)}} / \rho^{\textrm{(DE)}}$.

Definitions (\ref{DEd}) and (\ref{DEp}) lead to the continuity
equation for the dark energy component, namely,
\begin{align}\label{DEw}
\dot{\rho}^{\textrm{(DE)}}+3H(\rho^{\textrm{(DE)}}+p^{\textrm{(DE)}})=0.
\end{align}
Now, we can rewrite the effective equation of state in the following form
\begin{align}\label{weffd}
w^{\textrm{(eff)}}=\frac{F}{A}\Big{(}\Omega^{\textrm{(DE)}} w^{\textrm{(DE)}}+\frac{\Omega^{\textrm{(rad)}}}{3}\Big{)},
\end{align}
where we have defined
\begin{align}\label{Omrad}
\Omega^{\textrm{(rad)}}\equiv \frac{8\pi G \rho^{\textrm{(rad)}}}
{3H^2 F}~~~~~\mbox{and}~~~~~~\Omega^{\textrm{(DE)}}\equiv
\frac{8\pi G \rho^{\textrm{(DE)}}}{3H^2 F},
\end{align}
which lead to the usual density parameters for GR. Using definition (\ref{varx3}), $w^{\textrm{eff}}$ reads in a
suitable form
\begin{align}\label{weff2}
w^{\textrm{(eff)}}=\frac{1}{3} (1-2 x_{3}).
\end{align}

Also, for a general matter, the cosmological solutions, for a
constant value of $x_{3}$, can be found from equations
(\ref{varx3}) to be
\begin{align}\label{scalefactore}
a(t)=a_{0}\left(\frac{t-t_{i}}{t_{0}-t_{i}}\right)^{\frac{1}{2-x_{3}}}
\end{align}
and, for the conservation of the energy--momentum tensor one has
\begin{align}\label{densityequation}
\dot{\rho}(t)+2\left(2-x_{3}\right)H(t)\rho(t)=0,
\end{align}
where $a_{0}$ and $t_{0}$ are the integral constants that can be
fixed by the present values, and for $t_{i}$, we set $a(t_{i})=0$.
Equations (\ref{scalefactore}) and (\ref{densityequation}) hold
for all values of $x_{3}$ except for $x_{3}=2$. In this special
case we have $\dot{H}=0$, which leads to either a de~Sitter
solution or a static one.

In the next section, we consider a particular form of the function
$f(R,T)$ and show that the acceptable solution trajectories tend
to transit from the radiation era with $x_{3}=0$ to the dust--like
matter era with $x_{3}=1/2$, where, for these two values, the
conservation equation (\ref{densityequation})
gives\rlap,\footnote{The radiation and the dust--like types of
matters, analogously, are dictated from the appearance of the
corresponding equations.}\ respectively,
\begin{align}\label{third}
\dot{\rho}^{\textrm{(rad)}}+4H\rho^{\textrm{(rad)}}=0
\end{align}
and
\begin{align}\label{third1}
\dot{\rho}^{(m)}+3H\rho^{(m)}=0.
\end{align}
\section{Dynamical Systems Approach of the Minimal Case
         $f(R,T)=g(R)+h(T)$}\label{minimalsection}
In this section, we investigate the model, by employing the
dynamical systems approach, first when the geometrical sector and
the matter sector in the function $f(R,T)$ are minimally
coupled\rlap.\footnote{We apply the conventional terminology used
in the literature for adding and crossing two terms in the
Lagrangian as the minimal and the non--minimal couplings,
respectively.}\ The cases of non--minimal coupling are considered
in the later sections. In the minimal case, we assume that the
form of function $f(R,T)$ is
\begin{align}\label{form}
f(R,T)=g(R)+h(T),
\end{align}
where $h(T)$ and $g(R)$ are arbitrary functions and hereafter, we
show the functions $g(R)$, $h(T)$ and their derivatives without
indicating their arguments for the sake of convenience.

The dynamical systems approach\footnote{See Ref.~\cite{Wainwright}
and references therein.}\
 introduces a relatively simple technique
to investigate whole space of solutions in the form of some
extremum points (the fixed points), by which the evolution of
system can be pictured qualitatively near these points. A
qualitative study is possible via checking the phase space
trajectories, whose behaviors are sensitive to initial conditions.
In this way, one can obtain different descriptions dependent on
different initial conditions and therefore, indicates those
initial conditions that lead to a desired physical result. In
cosmological applications, by this technique, one is capable of
determining the early and the late time behaviors of models (in
addition to possible matter or radiation solutions). That is, one
can achieve a global picture of all solutions and behaviors of the
system near these solutions. As a result, through the dynamical
systems approach, the inconsistent models can be ruled out, and
also those models which deserve further investigation could be
selected. For a recent application of the dynamical systems
approach to some modified theories of gravity see, e.g.,
Refs.~\cite{FarSal10,FarSal11,HSS12}.

Now, rewriting equations (\ref{first}) and (\ref{second}) with
(\ref{form}) gives
\begin{align}\label{eom1}
1+\frac{g}{6H^{2} g'} +\frac{h}{6 H^{2} g'}-\frac{R}{6 H^{2}} + \frac{\dot{g'}}
{H g'}=\frac{8 \pi G \rho^{\textrm{(m)}}}{3H^{2} g'} +\frac{h' \rho^{\textrm{(m)}}}
{3H^{2} g'}+\frac{8 \pi G \rho^{\textrm{(rad)}}}{3H^{2} g'}
\end{align}
and
\begin{align}\label{eom2}
2\frac{\dot{H}}{H^{2}}+\frac{\ddot{g'}}{H^{2} g'} -\frac{\dot{g'}}{H g'}=-\frac{8
\pi G \rho^{\textrm{(m)}}}{H^{2} g'}-\frac{h' \rho^{\textrm{(m)}}}{H^{2} g'}
-\frac{32\pi G \rho^{\textrm{(rad)}}}{3H^{2} g'}.
\end{align}
In the approach of dynamical systems, original EOM (e.g.,
equations (\ref{eom1}) and (\ref{eom2}) in this work) can be cast
in the form of some new evolutionary EOM in terms of new variables
(which are constructed from the original ones) and their first
derivatives. Then, the solutions of these new EOM are indicated as
some fixed points of the system which are obtained through an
extremization, where if the new EOM do~not explicitly contain time
then, the system will be called an autonomous one. We employ this
approach to extract and analyze the solutions of equations
(\ref{eom1}) and (\ref{eom2}) by employing the introduced
variables (\ref{varx1})--(\ref{parameters}).

First of all, constraint (\ref{source}), for the minimal case with $h\neq\mbox{\textrm{constant}}$, gives
\begin{align}\label{deformed constraint1}
Th'' =-\frac{1}{2} h',
\end{align}
i.e., by (\ref{n}), $n=-1/2$, and by integrating with respect to the trace $T$, reads
\begin{align}\label{deformed constraint2}
Th' -\frac{1}{2} h +C=0,
\end{align}
where $C$ is an integration constant. This constant must be zero
to be consistent with condition (\ref{scons}), as we will show.
Thus, equation (\ref{deformed constraint2}) with $C=0$ leads to
$s=1/2$ and hence, the relation $x_{6}=x_{4}/2$. Therefore, with
these unique constants $n$ and $s$, the phase space variables of
the model are reduced from six to five. As we will see, this
reduction makes the problem to become more tractable than if there
is~not such a reduction.

Obviously, all cases with $x_{4}=0 $ (for a non--singular
denominator in (\ref{varx4})), in the minimal case, get returned
to $f(R)$ gravity, however, the cases with non--zero $x_{4}$ give
more general solutions than $f(R)$ gravity. Also, all cases with
$h=\textrm{constant}$ can be considered in $f(R)$ gravity
background and act as if they have a cosmological constant. Here,
by applying equation (\ref{deformed constraint1}), the only form
that respects the conservation law, in the minimal case, is
\begin{align}\label{the only minimal}
f(R,T)=g(R)+c_{1} \sqrt{-T}+c_{2},
\end{align}
where $c_{1}$ and $c_{2}$ are some constants with respect to $T$,
however, they, in general, can be functions of the Ricci scalar
$R$. Those cases in which $c_{1}$ is a function of $R$ will be
considered as a non--minimal case in the subsequent section. Now,
let us obtain the possible ``good'' cosmological solutions, i.e.
those solutions that describe a dust--like matter dominated era
followed by an accelerated era, for the general case
 (\ref{the only minimal}).

Equation (\ref{eom1}) gives a constraint that must hold for the
defined variables (\ref{varx1})--(\ref{varx5}) as
\begin{align}\label{minimal density}
\Omega^{\textrm{(m)}} \equiv \frac{8 \pi G \rho^{\textrm{(m)}}}{3 H^{2}g'}=1-x_{1}-x_{2}-x_{3}-x_{4}-x_{5}.
\end{align}
Hence, the autonomous EOM for the five independent variables
(\ref{varx1})--(\ref{varx5}) can be achieved via
\begin{align}
&\frac{d x_{1}}{d N}= -1+x_{1} (x_{1}-x_{3}) -3x_{2} -x_{3} -\frac{3}{2} {x_{4}}+x_{5} \label{minimal 1},\\
&\frac{d x_{2}}{d N}= \frac{x_{1} x_{3}}{m} +x_{2}\left(4+x_{1} -2x_{3}\right)\label{minimal 2},\\
&\frac{d x_{3}}{d N}=- \frac{x_{1} x_{3}}{m} +2x_{3} \left(2 -x_{3}\right)\label{minimal 3},\\
&\frac{d x_{4}}{d N}=x_{4}\left(\frac{5}{2}+x_{1}-2x_{3}\right)\label{minimal 4},\\
&\frac{d x_{5}}{d N}=x_{5}\left(x_{1}-2x_{3}\right)\label{minimal 5},
\end{align}
where $N$ represents derivative with respect to $\ln a$ and
equation (\ref{minimal density}) has been used. The solutions for
the system of equations (\ref{minimal 1})--(\ref{minimal 5}) for
arbitrary $m(r)$, $n(s)=-1/2$ and $s=1/2$ are listed in
Table~\ref{table:fixedpoints}. These solutions include ten fixed
points $P_{1}$--$P_{10}$ at which the variables $x_{1}$--$x_{5}$
(and any arbitrary function of them) take their critical values,
i.e. these solutions are those of the system of equations
$dx_{i}/dN=0,~i=1,\cdots,5$. Thus, in general, the parameters
$r=r(x_{2}, x_{3})$ and $s=s(x_{4}, x_{6})$ must take their
critical values too. That is,
\begin{align}\label{the only minimal1}
\frac{dr}{dN}=\frac{\partial r(x_{2}, x_{3})}{\partial x_{2}}\frac{dx_{2}}{dN}
+\frac{\partial r(x_{2}, x_{3})}{\partial x_{3}}\frac{dx_{3}}{dN}=0
\end{align}
and
\begin{align}\label{the only minimal2}
\frac{ds}{dN}=\frac{\partial s(x_{4}, x_{6})}{\partial x_{4}}\frac{dx_{4}}{dN}
+\frac{\partial s(x_{4}, x_{6})}{\partial x_{6}}\frac{dx_{6}}{dN}=0,
\end{align}
which, using definitions (\ref{varx2})--(\ref{varx4}) and (\ref{varx6})--(\ref{parameters}), give
\begin{align}\label{rcons}
0=\frac{d r}{d N}=-r\left(\frac{1+r+m(r)}{m(r)}\right) x_{1}\equiv-r {\mathcal M} (r) x_{1}
\end{align}
and
\begin{align}\label{scons}
0=\frac{d s}{d N}=3s\Big{(}s-n(s)-1\Big{)},
\end{align}
where we have defined
\begin{align}\label{bigM}
{\mathcal M} (r)\equiv \frac{1+r+m(r)}{m(r)},
\end{align}
which is well--defined for $m(r)\neq0$\,\rlap.\footnote{Note that,
all solutions that satisfy $m(r)=-r-1$ must satisfy ${\mathcal
M}(r)=0$ as a more strong constraint, this fact  affects the
analysis involved in Sec.~\ref{case studies}.}\ As a result, the
condition $ds/dN=0$ for $s\neq0$\rlap,\footnote{Note that, the
corresponding solutions with $s=0$ have been discarded, for they
contradict with the former result $n=-1/2$.}\ with $n=-1/2$, leads
to $s=1/2$ which in turn gives the constant $C$ in equation
(\ref{deformed constraint2}) to be zero. The acceptable solutions
are those that respect these two conditions $dr/dN$=0 and
$ds/dN=0$. Now, restoring constraint (\ref{source}), from
equations (\ref{rcons}) and (\ref{scons}), it turns out that all
acceptable solutions must lie in one of the following three
categories
\begin{align}\label{setss}
\left\{\begin{array}{l}
1)~r=0,~s=\frac{1}{2}=-n, \\
2)~{\mathcal M}(r)=0,~s=\frac{1}{2}=-n, \\
3)~x_{1}=0,~s=\frac{1}{2}=-n.
\end{array}\right.
\end{align}
\begin{table}[h]
\centering \caption{The fixed points solutions of the dynamical
systems approach of $f(R,T)=g(R)+h(T)$.}
\begin{tabular}{l l l l l l}\hline\hline

Fixed point     &Coordinates $(x_{1},x_{2},x_{3},x_{4},x_{5})$
&Scale factor
&$\Omega^{\textrm{(m)}}$    &$\Omega^{\textrm{(rad)}}$    &$w^{\textrm{(eff)}}$\\[0.5 ex]
\hline
$P_{1}$&$\left(\frac{3m}{2(1+m)},-\frac{5+8m}{4(1+m)^2},\frac{5+8m}{4(1+m)},\frac{4-m(3 + 10 m)}
{4(1+ m)^2},0\right)$& $a(t)=a_{0}\left(\frac{t-t_{i}}{t_{0}-t_{i}}\right)^{\frac{4(1+m)}{3}} $&$0$&$0$&$-\frac{1+2m}{2(1+m)}$\\[0.5 ex]
$P_{2}$&$\left(\frac{2(1- m)}{1 + 2m},\frac{1-4 m}{m (1+2m)},-\frac{(1-4m)(1+m)}{m(1+2m)},0,0\right)$
&$a(t)=a_{0}\left(\frac{t-t_{i}}{t_{0}-t_{i}}\right)^{\frac{m(1+2 m)}{1-m}}$&$0$&$0$&$\frac{2-5m-6m^2}{3m(1+2m)}$\\[0.5 ex]
$P_{3}$&$\left(\frac{3 m}{1 + m},~-\frac{1 + 4 m}{2 (1 + m)^2},\frac{1+4m}{2(1+m)},0,0\right)$&$a(t)
=a_{0}\left(\frac{t-t_{i}}{t_{0}-t_{i}}\right)^{\frac{2(1+m)}{3}}$&$\frac{2-m(3+8m)}{2(1+m)^2}$&$0$&$-\frac{m}{1+m}$\\[0.5 ex]
$P_{4}$ & $\left(-4,5,0,0,0\right)$&$a(t)=a_{0}\left(\frac{t-t_{i}}{t_{0}-t_{i}}\right)^{\frac{1}{2}}$&$0$&$0$&$\frac{1}{3}$\\[0.5 ex]
$P_{5}$ & $\left(-\frac{5}{2},0,0,~\frac{7}{2},0\right)$&$a(t)=a_{0}\left(\frac{t-t_{i}}{t_{0}-t_{i}}
\right)^{\frac{1}{2}}$&$0$&$0$&$\frac{1}{3}$\\[0.5 ex]
$P_{6}$ & $\left(-1,0,0,0,0\right)$&$a(t)=a_{0}\left(\frac{t-t_{i}}{t_{0}-t_{i}}\right)^{\frac{1}{2}}$&$2$&$0$&$\frac{1}{3}$\\[0.5 ex]
$P_{7}$ & $\left(1,0,0,0,0\right)$&$a(t)=a_{0}\left(\frac{t-t_{i}}{t_{0}-t_{i}}\right)^{\frac{1}{2}}$&$0$&$0$&$\frac{1}{3}$\\[0.5 ex]
$P_{8}$\footnote{This solution has $\dot{H}=0.$} & $\left(0,-1,2,0,0\right)$&$a(t)=a_{0}\exp{H_{0}t}$&$0$&$0$&$-1$\\[0.5 ex]
$P_{9}$ & $\left(0,0,0,0,1\right)$&$a(t)=a_{0}\left(\frac{t-t_{i}}{t_{0}-t_{i}}\right)^{\frac{1}{2}}$&$0$&1&$\frac{1}{3}$\\[0.5 ex]
$P_{10}$ & $\left(\frac{4m}{1+4m},-\frac{2m}{(1+m)^2},\frac{2m}{1+m},0,\frac{1-m(2+5m)}{(1+m)^2}
\right)$&$a(t)=a_{0}\left(\frac{t-t_{i}}{t_{0}-t_{i}}\right)^{\frac{1 + m}{2}}$&$0$ &$\frac{1-m(2+5m)}{(1+m)^2}$&$\frac{1-3m}{3(1+m)}$\\
 \hline\hline
\end{tabular}
\label{table:fixedpoints}
\end{table}

A glance at Table~\ref{table:fixedpoints} shows that the points
$P_{1}$,~$P_{2}$,~$P_{3}$ and $P_{10}$ satisfy the condition
$m(r)=-r-1$, the parameter $r$ vanishes for the point $P_{4}$, and
for the points $P_{8}$ and $P_{9}$, we have $x_{1}=0$. These are
the only obvious points that respect (\ref{setss}). The other
points have both $x_{2}=0$ and $x_{3}=0$, which implies that there
is an ambiguity in determining $r$ clearly. Nevertheless, these
points actually do satisfy equations (\ref{rcons}) and
(\ref{scons}) and hence, $r$ can be determined by a
straightforward calculation using definition (\ref{parameterr}).
In this respect, we consider that the condition $m(r)=-r-1$ should
be valid for all of the points, and use it wherever it is
necessary.

In the following discussions, the stability analysis of the fixed
points are performed via inspecting the corresponding eigenvalues
of them. Imprecisely speaking, the trajectories of the phase space
advance to a fixed point if all eigenvalues have negative values,
and recede from a fixed point if all eigenvalues have positive
values. In this respect, the fixed points occurring in the former
and the latter sets are called the stable and unstable points,
respectively. The fixed points with both positive and negative
eigenvalues are called saddle points, and those trajectories which
advance to a saddle fixed point along some eigenvectors may recede
from it along some other eigenvectors.

In Subsection~\ref{matteronly}, we investigate the properties of
each of the fixed points of Table~\ref{table:fixedpoints} in the absence of the radiation. Since
the calculations in a system with five degrees of freedom can be
very messy and time consuming to study, we consider the effects of
the radiation in Subsection~\ref{radiation}. Also, in
Subsection~\ref{sub3}, we illustrate ``good'' cosmological
solutions, i.e. those solutions that determine the trajectories
which connect the dust--matter dominated points to the accelerated
expansion dominated points. Incidentally, the considerations have
been assisted by numerical manipulations wherever the exact
computations have~not been possible.
\subsection{Properties of Fixed Points in the Absence of Radiation}\label{matteronly}
In the absence of radiation, there are only the first eight
fixed points $P_{1}$--$P_{8}$. While presenting the properties of
these points, we compare the results with the corresponding
results of the $f(R)$ gravity in Ref.~\cite{amend} (whenever it is
necessary), and briefly indicate the obtained results in
Table~\ref{table:table2}.
\begin{itemize}\label{item0}
\item The Point $P_{1}$

This is a new fixed point which corresponds to a
curvature--dominated point\rlap.\footnote{We refer to a point with
both properties $\Omega^{\textrm{(m)}}=0=\Omega^{\textrm{(rad)}}$
as a curvature--dominated point.}\ This point can play the role of
an accelerated expansion point provided that
$w^{\textrm{(eff)}}<-1/3$ for $m>-1/4$ and $m<-1$. In the former
range, we have a non--phantom accelerated universe, and the latter
one lies in a phantom domain. The eigenvalues of this point are
obtained as
\begin{align}\label{eig1}
-\frac{3}{2},~~~~~-\frac{3m(1+m)(3+2m)+a(m)}{8m(1+m)^2},~~~~~\frac{-3m(1+m)(3+2m)+a(m)}{8m(1+m)^2},~~~~~\frac{3}{2}(1+m'),
\end{align}
where
\begin{align}\label{am}
a(m)\equiv \Bigg{\{}m (1 + m)^2 \Big{[}-160 + m (-55 + 700 m + 676 m^2)\Big{]}\Bigg{\}}^{1/2},
\end{align}
and $m'\equiv dm/dr$. The above eigenvalues show that with
$m'>-1$, we have a saddle point. However, for $m'<-1$, the point
$P_{1}$ is a stable point when $-4/5<m<-5/8$ or $0.43<m<1/2$ with
real valued eigenvalues, a spiral stable point when $0<m\leq0.43$
and a saddle point otherwise. Nevertheless, within these ranges,
the first one does~not lead to the condition
$w^{\textrm{(eff)}}<-1/3$ and hence, we discard it. As a result,
the point $P_{1}$ introduces two new ranges that can accelerate
universe in the non--phantom domain, which collectively are
\begin{align}\label{p1-dom}
m'<-1,~~~~~~0<m<\frac{1}{2},~~~~~-\frac{2}{3}<w^{\textrm{(eff)}}<-\frac{1}{2}.
\end{align}

In the limit $\mid m \mid \rightarrow0$, the eigenvalues take the values
\begin{align}\label{eig1Lim}
-\frac{3}{2},~~~~~-\frac{9}{8}+\sqrt{-\frac{5}{2m}},~~~~~-\frac{9}{8}-\sqrt{-\frac{5}{2m}},~~~~~\frac{3}{2} (1 +m').
\end{align}
It means that, for $m\rightarrow0^{+}$, this point is a spiral
stable point when $m'<-1$ and a saddle one otherwise. When $\mid m
\mid \rightarrow \infty$, the point tends to a de~Sitter point
with coordinates $(3/2,~0,~2,~-5/2$), which is~not a stable point.
Indeed, from (\ref{eig1}), it is obvious that $P_{1}$ is a
permanent saddle point in the both limits $m'=0$ and $\mid m \mid
\rightarrow \infty$.
\end{itemize}
\begin{itemize}
\item The Point $P_{2}$

The point $P_{2}$ has also
$\Omega^{\textrm{(m)}}=0=\Omega^{\textrm{(rad)}}$, and like
$P_{1}$ is a curvature--dominated point whose effective
equation of state depends on the parameter $m$. An accelerated
expansion behavior can be achieved when $m <(-1-\sqrt{3})/2$ or
$(-1 + \sqrt{3})/2 < m < 1$ in the non--phantom domain, and when
$-1/2 < m < 0$ or $m >1$ in the phantom domain. The eigenvalues
are obtained to be
\begin{align}\label{eig2}
-4+\frac{1}{m},~~~~~\frac{-8m^{2}-3m+2}{m(1+2m)},~~~~~\frac{2 (1 - m^2)(1 +m')}{m(1 + 2 m)},~~~~~\frac{-10m^2-3m+4}{2m(1+2m)}.
\end{align}
In the limit $\mid m \mid \rightarrow \infty$, it asymptotically
reaches the point $P_{2,{\rm dS}}=(-1,~0,~2,~0)$, at which
universe expands as a de~Sitter accelerated one, and is a stable
point for $m'>-1$. In the opposite limit, when $\mid m \mid
\rightarrow0$, the eigenvalues take the following forms
\begin{eqnarray*}\label{eig2Lim}
\frac{1}{m},~~~\frac{2}{m},~~~\frac{2}{m}(1+m'),~~~\frac{2}{m}.
\end{eqnarray*}
Thus, in order to have a stable acceleration era, one must have $
m \rightarrow 0^{-}$ and $ m'>-1$, simultaneously. An
investigation of the eigenvalues gives ranges of $m$, in which one
can expect a stable accelerated expansion behavior. For the
non--phantom domain, we have
\begin{align}\label{non-ph}
&\mathcal{A})~~m'>-1,~~~~~~m<-\frac{1}{2}(1+\sqrt{3}),~~~~~~-1<w^{\textrm{(eff)}}<-\frac{1}{3},\\
&\mathcal{B})~~m'<-1,~~~~~~~\frac{1}{2}<m<1,~~~~~~~~~~~~~~-1<w^{\textrm{(eff)}}<-\frac{2}{3},
\end{align}
and for the phantom domain, we obtain
\begin{align}\label{ph}
&\mathcal{C})~~m'>-1,~~~~~~~m>1,~~~~~~~~~~~~~~~~~~-1.07<w^{\textrm{(eff)}}<-1,\\
&\mathcal{D})~~m'>-1,~~~~~~-\frac{1}{2}<m<0,~~~~~~~~~~w^{\textrm{(eff)}}<-7.60.
\end{align}

However, $P_{2}$ is an unstable point in the range $0 < m < 1/4$
provided that $m'> -1$. The properties of this point do~not change
in this model compared to the $f(R)$ gravity, except in the case
$\mathcal{B}$, where the range $m$ becomes more restricted, i.e.
the corresponding range is $(\sqrt{3}-1)/2<m<1$ in the case
$\mathcal{B}$ in the $f(R)$ gravity.
\end{itemize}
\begin{itemize}
\item The Point $P_{3}$

The point through which we can search for a matter era is $P_{3}$,
which also appears in the $f(R)$ gravity. For $m=0$, we have
$w^{\textrm{(eff)}}=0$ and $\Omega^{\textrm{(m)}}=1$. The
eigenvalues are
\begin{align}\label{eig3}
\frac{3}{2},~~~~\frac{-3m+b(m)}{4m(1+m)},~~~~\frac{-3m-b(m)}{4m(1+m)},~~~~3(1+m'),
\end{align}
where
\begin{align}
b(m)\equiv \Big{[}m\big{(}256 m^3+160 m^2 - 31 m-16\big{)}\Big{]}^{1/2}.
\end{align}
The existence of the positive constant eigenvalue $3/2$ makes the
point $P_{3}$ not to be a stable point, instead, it is always a
saddle point. It is an interesting result that does~not occur in
the $f(R)$ gravity. For infinitesimal values of the parameter $m$,
we can approximate the eigenvalues as
\begin{align}\label{eigenvalues-app3}
\frac{3}{2},~~~~-\frac{3}{4}+\sqrt{-\frac{1}{m}},~~~~-\frac{3}{4}-\sqrt{-\frac{1}{m}},~~~~3(1+m').
\end{align}
In the limit $m\rightarrow0^{+}$, we have an acceptable saddle
point matter era. However, the point $P_{3}$, in the limit
$m\rightarrow 0^{-} $ is~not generally acceptable, for the second
eigenvalue becomes a large positive real value. Therefore, the
matter era becomes very short, so that the observational data
cannot be matched. The point $P_{3}$ contains some ranges in which
universe can be accelerated, but not in a usual way, for, the
accelerating conditions are
\begin{align}\label{accelerating cond. p3}
& \mathcal{E}) ~m>\frac{1}{2},~~~~~~~~-1 < w^{(\textrm{eff})} < -\frac{1}{3},~~~~~-4<\Omega^{(m)}<-\frac{1}{3},\\
& \mathcal{F}) ~m<-1,~~~~~~~ w^{(\textrm{eff})} <
-1,~~~~~~~~~~~~~\Omega^{(m)}<-4.
\end{align}
That is, the accelerated expansion can occur with a negative value
for the matter density parameter, which is~not physically
interesting. Considering the definition used in relation
(\ref{minimal density}), the solutions denoting
$\Omega^{\textrm{(m)}}<0$ are ruled out in the background of
viable $f(R)$ models with the condition\footnote{This condition
guaranties that the gravity force is an attractive one. As $f(R)$
theories are special cases with $h(T)=0$ in the minimal coupling
case, hence, this condition should hold.} $g'(R)>0$, which we have
also adopted here.
\end{itemize}
\begin{itemize}
\item The Points $P_{4}$,~$P_{5}$~and~$P_{7}$

There are three points in $f(R,T)$ gravity with
$\Omega^{\textrm{(m)}}=0$ and $\Omega^{\textrm{(rad)}}=0$ whose
equations of state mimic the one for radiation. As these points
do~not correspond to any known matter, they are~not physically
interesting. Hence, we discard these solutions in
Subsection~\ref{sub3} as non--physical ones.

The point $P_{4}$ is a special case of $P_{2}$ if $m$ is set to be $m=-1$, and its eigenvalues are found to be
\begin{align}\label{eigenvalues4}
-5,~~~-3,~~~4(1 + \frac{1}{m}),~~~-\frac{3}{2}.
\end{align}
When $-1<m<0$, the point $P_{4}$ is a stable point, and is a
saddle one otherwise. This property has the same features in the
$f(R)$ gravity.

The point $P_{5}$ is a new solution, which does~not appear in the $f(R)$ gravity. The eigenvalues are derived as
\begin{align}\label{eigenvalues5}
-\frac{7}{2},~~-\frac{3}{2},~~\frac{m(5+11m)-5r(1+r)m'-5c(m,m')}{4m^2},~~\frac{m(5+11m)-5r(1+r)m'+5c(m,m')}{4m^2},
\end{align}
where
\begin{align}\label{cm}
c(m,m')\equiv \Bigg{\{}m^2(1+m)^2+rm'\Big{[}-2m(1+m)+2(-1+m)mr+r(1+r)^2 m'\Big{]}\Bigg{\}}^{1/2}.
\end{align}
As it is obvious, the point $P_{5}$ never becomes unstable. When
$m$ is a non--zero constant, we have the eigenvalues as
\begin{align}\label{eigenvalues5-constant m}
-\frac{7}{2},~~~-\frac{3}{2},~~~\frac{3}{2},~~~4+\frac{5}{2m},
\end{align}
i.e., $P_{5}$ is a saddle point for constant $m$. When $m\rightarrow0$, the eigenvalues become
\begin{align}\label{eigenvalues-app5}
-\frac{7}{2},~~~-\frac{3}{2},~~~-\frac{5m'}{2m},~~~\frac{5}{2m}.
\end{align}
Therefore, when $m\rightarrow0^{-}$ with $m'<0$, this point is a stable one, otherwise is a saddle one.

The last point in this category is $P_{7}$, which is regarded as a
special case of the point $P_{2}$ for $m=1/4$. This point has
eigenvalues
\begin{align}\label{eigenvaluesP7}
\frac{7}{2},~~~2,~~~\frac{m(-1+9m)+r(1+r)m'-c(m,m')}{2m^2},~~~\frac{m(-1+9m)+r(1+r)m'+c(m,m')}{2m^2}.
\end{align}
Thus, $P_{7}$ cannot be a stable point. When $m$ is a non--zero
constant, eigenvalues (\ref{eigenvaluesP7}) read
\begin{align}\label{eigenvaluesP7-constant m}
\frac{7}{2},~~~2,~~~4-\frac{1}{m},~~~5,
\end{align}
i.e., for $0<m<1/4$ , the point $P_{7}$ is a saddle point, and otherwise, it
behaves as an unstable point. In the limit $m\rightarrow0$, the
eigenvalues behave as
\begin{align}\label{eigenvalues-app7}
\frac{7}{2},~~~2,~~~-\frac{1}{m},~~~\frac{m'}{m},
\end{align}
where for $m\rightarrow0^{-}$ and $m'<0$, this point is unstable, and otherwise, is a saddle point.
\end{itemize}
\begin{itemize}
\item The Point $P_{6}$

This is a point with an unusual feature. The value of the density
parameter $\Omega^{\textrm{(m)}}$ does~not match the equation of
state in a meaningful manner, for we have $w^{\textrm{(eff)}}=1/3$
and $\Omega^{\textrm{(m)}}=2$. However, in this model, in the
evolution of universe and dependent on the stability of this
point, it may occur that universe approaches to this point. Hence,
like the points $P_{4}$, $P_{5}$ and $P_{7}$, the stability of
this point should be considered. The eigenvalues are given by
\begin{align}\label{eigenvalues6}
-2,~~~\frac{3}{2},~~~\frac{m(1+7m)-r(1+r)m'-c(m,m')}{2 m^2},~~~\frac{m(1+7m)-r(1+r)m'+c(m,m')}{2 m^2}.
\end{align}
The first two eigenvalues, $-2$ and $3/2$, show that this point is always a saddle one for all values of $m$ and $m'$.
\end{itemize}
\begin{itemize}
\item The Point $P_{8}$

The point $P_{8}$ is the only de~Sitter fixed point of
minimally--coupled form of $f(R,T)$ gravity. The corresponding
eigenvalues are represented as
\begin{align}\label{eigenvalues8}
-3,~~~-\frac{3}{2},~~~\frac{1}{2}\Big{(}-3-\sqrt{25-\frac{16}{m}}~\Big{)},~~~\frac{1}{2}\Big{(}-3+\sqrt{25-\frac{16}{m}}~\Big{)}.
\end{align}
This point is a stable one in the range $0<m<1$, and otherwise, is a saddle point.
\end{itemize}
\begin{table}[h]
\centering
\caption{The stability of the fixed points in both $f(R,T)$ and $f(R)$ gravities without radiation. }
\begin{tabular}{l@{\hskip 0.2in}ll}\hline\hline
\\
Fixed point     &Stability in $f(R,T)$ gravity      &Stability in the $f(R)$ gravity \\[2 ex]
\hline
\\
$P_{1}$&$\left\{\begin{array}{l}
a)~\forall~m,~m'>-1,~\textrm{Saddle} \\b)~0<m<1/2,~m'<-1,~\textrm{Stable} \\c)~m\rightarrow
\pm \infty,~\forall~m',~\textrm{Saddle} \\\end{array}\right.$& \textrm{Does~not appear}\\[0.5 ex]
\\
$P_{2}$&$\left\{\begin{array}{l}
d)~ 0<m<1/4,~m'>-1,~\textrm{Unstable}\\ e)~m<-1/2(1+\sqrt{3}),~m'>-1,~\textrm{Stable}\\ f)~-1/2<m<0,~m'>-1,~\textrm{Stable} \\
 g)~m>1,~m'>-1,~\textrm{Stable}\\h)~1/2<m<1,~m'<-1,~\textrm{Stable}\\i)~m\rightarrow \pm \infty,~m'>-1,~\textrm{Stable}\\
  \end{array}\right.$&$\left\{\begin{array}{l}
\textrm{The same properties except for}\\ h)~(1/2)(-1+\sqrt{3})<m<1,~m'<-1,~\textrm{Stable} \\ \end{array}\right.$\\[0.5 ex]
\\
$P_{3}$&\textrm{Always is a Saddle point}&$\left\{\begin{array}{l}
j)~0<m<0.327,~m'>-1,~\textrm{Saddle} \\k)~\forall~m,~m'=0,~\textrm{Saddle} \\ \end{array}\right.$\\[0.5 ex]
\\
$P_{4}$&$\left\{\begin{array}{l}
l)-1<m<0,~\textrm{Stable; otherwise Saddle}\\ m)~P_{4}=P_{2~\mid m=-1}\\\end{array}\right.$&~\textrm{The same properties}\\[0.5 ex]
\\
$P_{5}$&$\left\{\begin{array}{l} n)~\forall
m,~m'=0,~\textrm{Saddle; otherwise Saddle or
Stable}\\o)~m\rightarrow0^{-},~m'<0,~\textrm{
Stable};~\textrm{otherwise Saddle}\\\end{array}\right.$&\textrm{Does~not appear}\\[0.5 ex]
\\
$P_{6}$&~\textrm{Always is a Saddle point}&$p)\forall~m,~m'=0,$~\textrm{Saddle; otherwise Saddle or Stable}\\[0.5 ex]
\\
$P_{7}$&$\left\{\begin{array}{l} q)~\forall~m,
m'\neq0,~\textrm{Saddle or Unstable}\\
r)~0<m<1/4,~m'=0,~\textrm{Saddle};~\textrm{otherwise
Stable}\\ s)~m\rightarrow0^{-}, m'<0,~\textrm{Unstable};~\textrm{otherwise Saddle}\\t)~P_{7}=P_{2~\mid m=1/4}\\
\end{array}\right.$&~\textrm{The same properties}\\[0.5 ex]
\\
$P_{8}$&$\left\{\begin{array}{l}
u)~0<m<16/25,~r=-2,~\textrm{Spiral Stable}\\v)~16/25\leq m<1,~r=-2,~\textrm{Stable}\\w)~\textrm{Otherwise
Saddle}\\\end{array}\right.$&\textrm{The same properties}\\[0.5 ex]
\hline
\end{tabular}
\label{table:table2}
\end{table}
\subsection{Effects of Radiation}\label{radiation}
In this subsection, we take into account the effects of radiation
for the fixed points, and particularly check any possible change
in the stability of the fixed points\footnote{As discussed before,
the points $P_{4}$--$P_{7}$ do~not have physical meaning and hence, we
do~not consider them in this subsection.} $P_{1},~P_{2},~P_{3}$
and $P_{8}$.

The existence of radiation adds two new fixed points $P_{9}$ and
$P_{10}$, as shown in Table~\ref{table:fixedpoints}. The point
$P_{9}$ is a standard radiation point with the eigenvalues
$(4,4,5/2,-1,1)$, which denotes that this point is always a saddle
point the same as in the $f(R)$ gravity.

The eigenvalues of $P_{10}$ are given as
\begin{align}
\frac{5}{2},~~~1,~~~\frac{m-1+\sqrt{81m^2+30m-15}}{2(m+1)},~~~\frac{m-1-\sqrt{81m^2+30m-15}}{2(m+1)},~~~4(1+m').
\end{align}
It is interesting that the point $P_{10}$ is always a saddle one
irrespective of the values of $m$ and $m'$, for, numerically, it
is impossible that the third and forth eigenvalues simultaneously
take positive values. Furthermore, $P_{10}$, in the limit
$m\rightarrow0$, gives another radiation fixed point, in which the
eigenvalues are non--singular, i.e. they are achieved as
\begin{align}
\frac{5}{2},~~~1,~~~\frac{-1+i\sqrt{15}}{2},~~~\frac{-1-i\sqrt{15}}{2},~~~4(1+m'),
\end{align}
where $m'$ must be evaluated at $r\rightarrow-1$.

The inclusion of radiation does~not change the stability
properties of the eigenvalues of the other fixed points. In fact,
the addition of radiation to the action leads to the appearance of
the values $-5/2$, $(2 - 4 m - 10 m^2)/[m (1 + 2 m)]$, $-1$ and
$-4$ as the fifth eigenvalues of $P_{1}$, $P_{2}$, $P_{3}$ and
$P_{8}$, respectively. Thus, it is obvious that none of the
stability properties of the accelerated fixed points and the
matter point $P_{3}$ do change. This means that all the
cosmological solutions which have a true sequence
$P_{3}\rightarrow P_{1,2,8}$ can include a saddle radiation era
for $m\rightarrow0^{+}$.
\subsection{Cosmological Solutions}\label{sub3}
``Good'' cosmological solutions are those that pass a long enough
matter dominated era followed by an accelerated expansion, hence,
any matter point contained in the model must be a saddle point in
the phase space. However, the eras that show the accelerated
expansion should be an attractor (a stable point) in the phase
space. In this study, the only point that involves a matter point
is $P_{3}$ for $\mid m \mid \rightarrow 0^{+}$, and $P_{1}$,
$P_{2}$, $P_{3}$ and $P_{8}$ can be the accelerated points.
Hereafter, we indicate the matter point $P_{3}$ with the condition
$m(r\approx-1)\rightarrow0$ as $P^{(0)}_{3}$. It is worth
mentioning that any well--defined curve $m(r)$ of each model must
satisfy the relations $m(r_{i})=-r_{i}-1$ and ${\mathcal
M}(r_{i})=0$ for some root $r_{i}$, the second condition is equal
to the first one for cases containing $m(r_{i})\neq0$. The
equation $m(r_{i})=-r_{i}-1$ gives some roots that belong to the
points $P_{1}$, $P_{2}$ or $P_{3}$ which we generally indicate as
$P_{1(a, b,\cdots)}$, $P_{2(a, b,\cdots)}$ or $P_{3(a,
b,\cdots)}$.

\begin{figure}[h]
\epsfig{figure=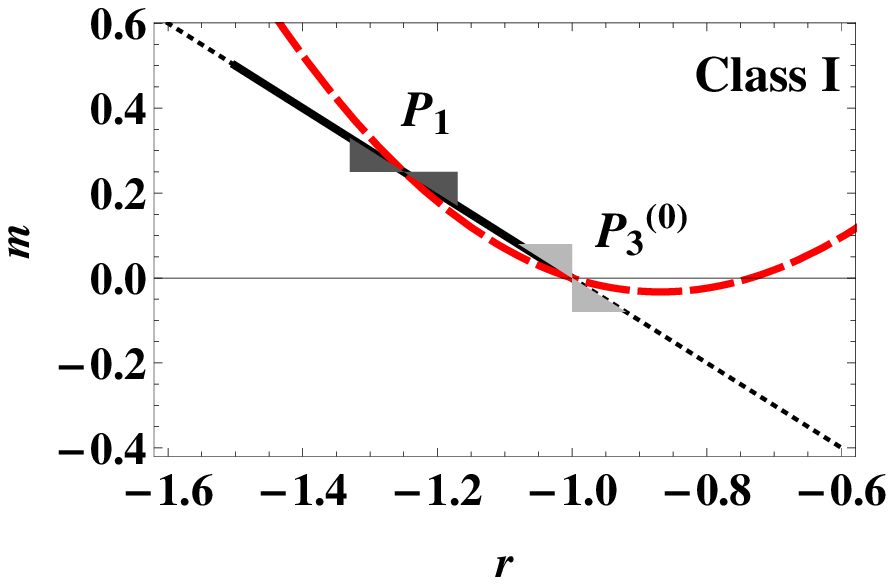,width=7cm}\hspace{4mm}
\epsfig{figure=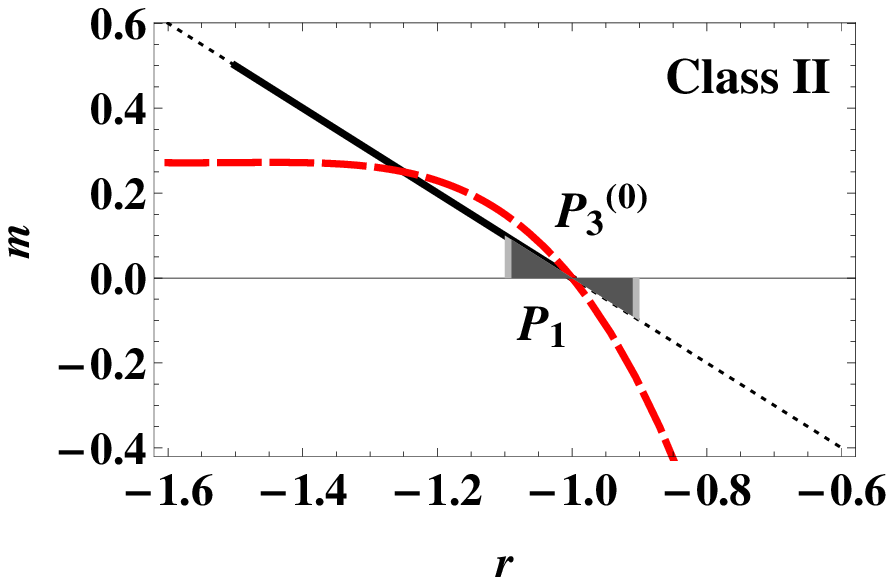,width=7cm}\vspace{4mm}
\epsfig{figure=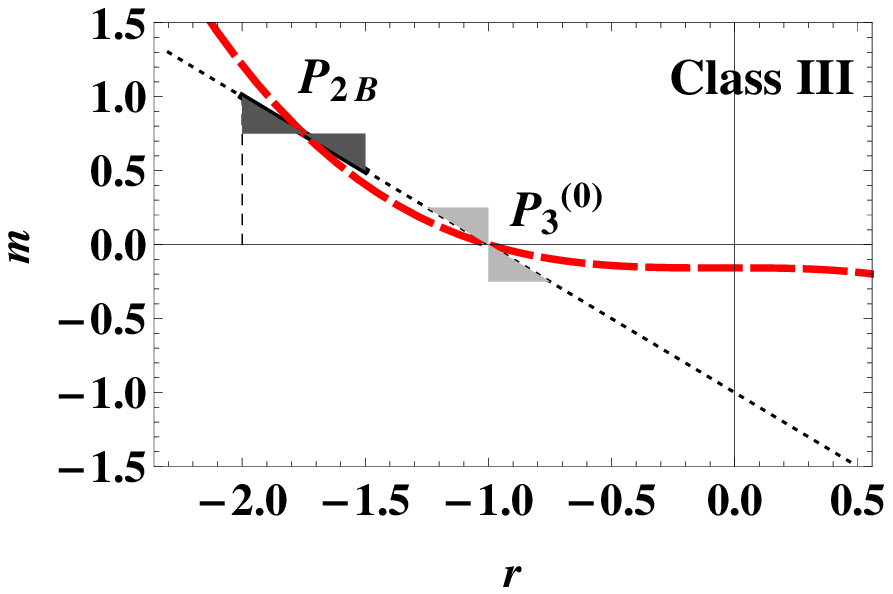,width=7cm}\hspace{4mm}
\epsfig{figure=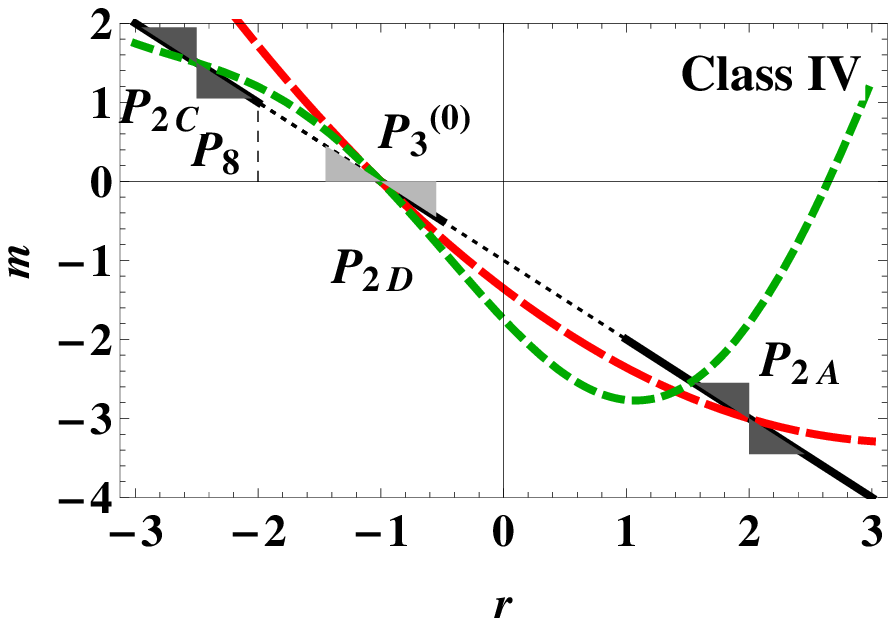,width=7cm}\vspace{4mm}
\epsfig{figure=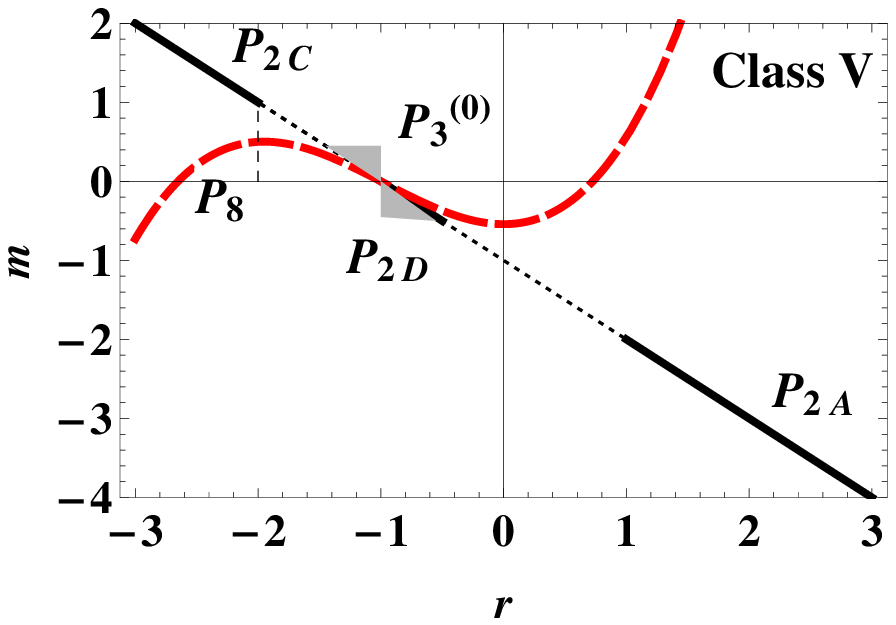,width=7cm}\hspace{4mm}
\epsfig{figure=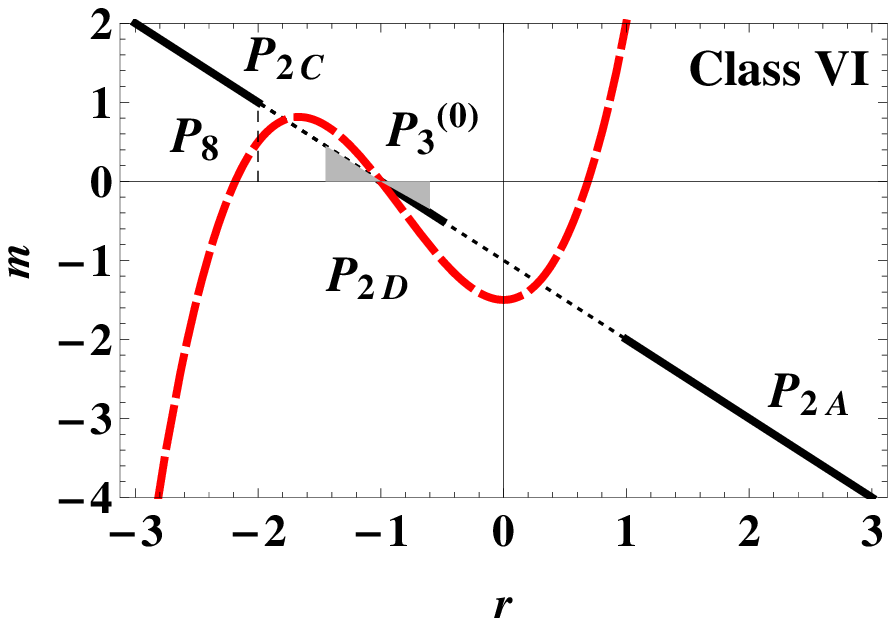,width=7cm} \caption{(color online).
\footnotesize {\textbf{Acceptable cosmological solutions of
$f(R,T)=g(R)+h(T)$ gravity.}} The classification of $f(R,T)$ model
in the $(r,m)$--plane. The line $m=-r-1$ and different curves of
$m(r)$, for the six classes of acceptable cosmological solutions,
are plotted. The transitions are depicted from the matter epoch
$P^{(0)}_{3}$ to the accelerated point $P_{1}$ in Classes~$I$
and~$II$, to the accelerated point $P_{2}$ in Classes~$III$
and~$IV$, and to the de~Sitter point $P_{8}$ in Classes~$V$
and~$VI$. The matter--acceleration epoch transition occurs in
Class~$II$ for the same value of $r$, and in Class~$VI$, before
reaching to the de~Sitter point $P_{8}$, with a non--stable
acceleration middle stage. The solutions are permitted only in the
black solid regions on the line $m=-r-1$ provided that
$m'_{1}<-1$, and $m'_{2,\mathcal{A},\mathcal {D},\mathcal{C}}>-1$.
For $P_{3}$, we can have either $m'_{3}<-1$ or $m'_{3}>-1$
depending on the corresponding class. In Classes $I$, $II$ and
$V$, we have $m'_{3}>-1$ whilst in the rest, we have $m'_{3}<-1$.
Unallowable slopes for the curve $m(r)$ are indicated by the light
gray triangles for $P^{(0)}_{3}$ (actually, there is no
unallowable slope for $P^{(0)}_{3}$, however, the light gray
triangles are indicated for the sake of classification) and by the
gray ones for $P_{1}$ and $P_{2}$. The dashed curves show
hypothetical curves which intersect the line $m=-r-1$ in the
critical points $P_{1}$, $P_{3}$ and $P_{2}$ in the regions
$\mathcal{A}$, $\mathcal{D}$ and $\mathcal{C}$. All of the classes
of solutions are new ones in $f(R,T)$ gravity except for Classes
$III$ and $V$ which also appear in the $f(R)$ gravity.}
\label{Figa}
\end{figure}

\begin{figure}[h]
\epsfig{figure=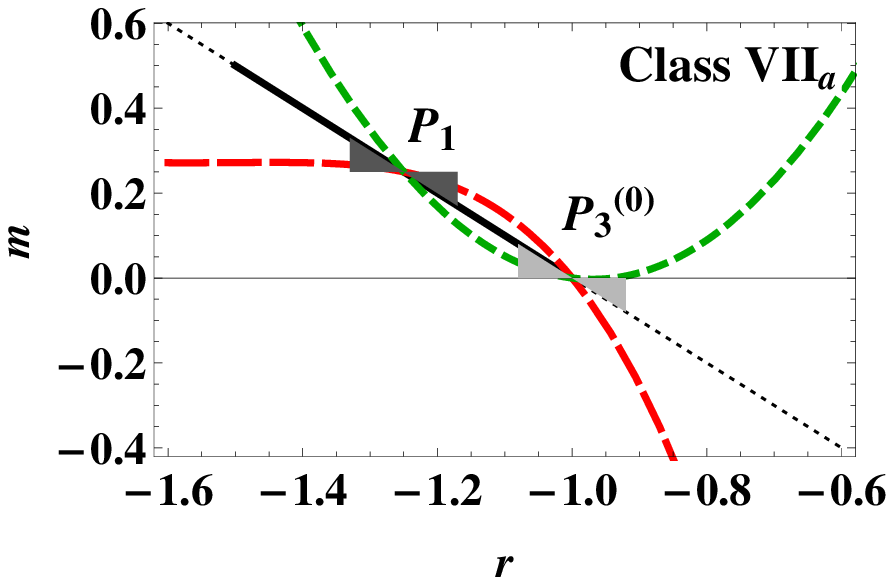,width=7cm}\hspace{4mm}
\epsfig{figure=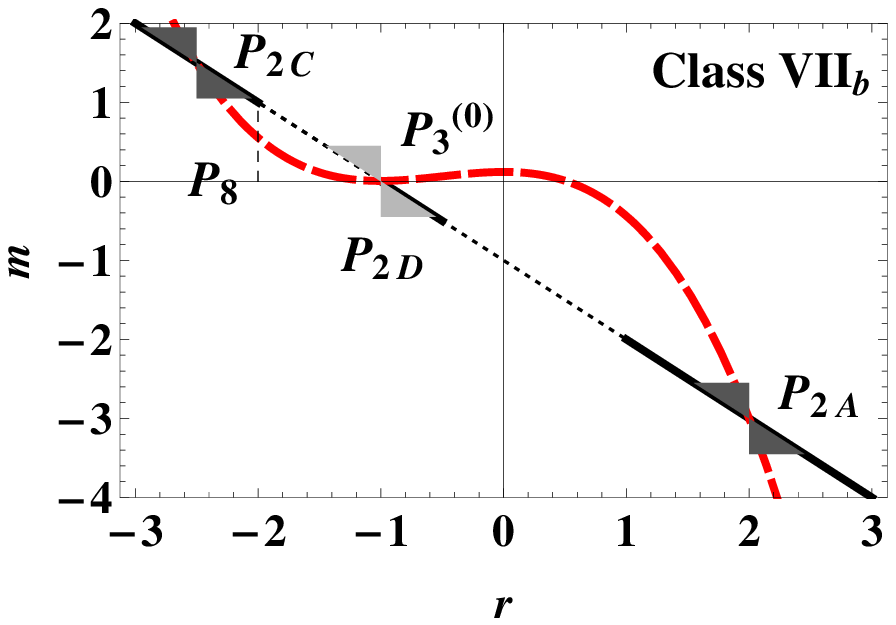,width=7cm}\vspace{4mm}
\epsfig{figure=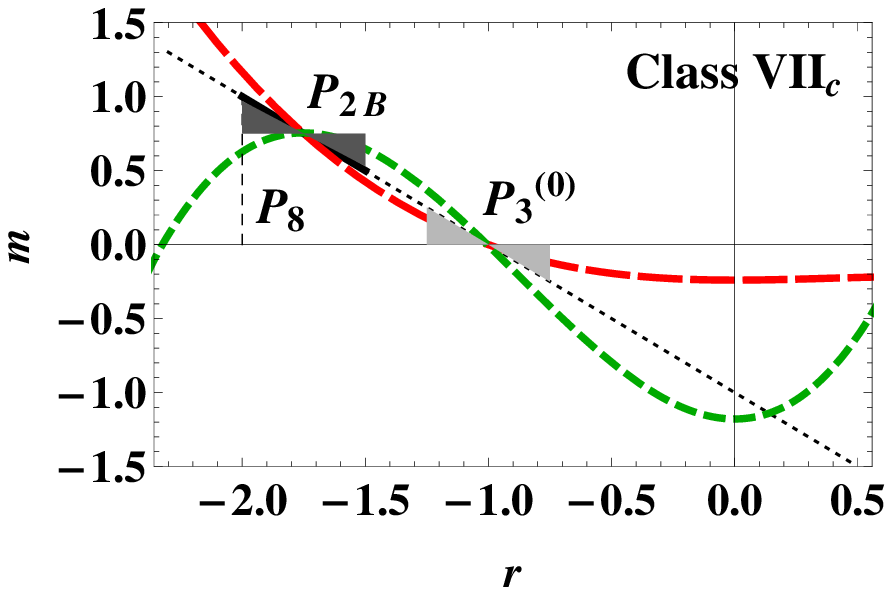,width=7cm}\ \caption{(color online).
\footnotesize {\textbf{Unacceptable cosmological solutions of
$f(R,T)=g(R)+h(T)$ gravity.}} Some classes of solutions, that
suffer from either the absence of a matter dominated epoch or a
stable accelerated era or unallowed transitions from matter to
acceleration phase, are presented. There may be some other classes
related to these solutions, but all of them can be accounted as
subclasses of the mentioned ones. Again, as in Figure~\ref{Figa},
the dashed curves show hypothetical curves which intersect the
line $m=-r-1$.} \label{Figb}
\end{figure}

The accelerating roots of $P_{3}$ labeled by $a, b,\cdots$
correspond to some negative matter density parameters as are shown
in Subsection~\ref{matteronly} and therefore, cannot be a physical
one, hence, we discard them. Consequently, we should consider the
cosmological transitions of $P^{(0)}_{3}$ to either $P_{1}$,
$P_{2}$ or $P_{8}$. Another assumption that we apply in the rest
of this work is to discard solutions with $m \rightarrow 0^{-}$,
for from (\ref{eigenvalues-app3}), it is obvious that one of the
eigenvalues gets a large positive value for small negative value
of $m$ and hence, diverges for infinitesimal negative values. This
means that, the trajectories leave the matter era very fast and
hence, the matter era becomes very short which causes difficulties
in matching the model with the observations. Thus, in general, the
models with $m \rightarrow 0^{-}$ are unacceptable. We indicate in
the following classification that, as $P_{3}$ is a saddle point
irrespective of the values of $m$ and $m'$, there are more
cosmological solutions than in the $f(R)$ gravity. We study these
transitions in turn, and suppose that there are some roots in all
important regions for the generality of the discussion. Also, we
assume that the condition $m(r_{i})=-r_{i}-1$ holds with
$m(r_{i})\neq0$.
\begin{itemize}\label{item1}
\item $P_{3}(m'_{3}>-1, m>0)$~\textrm{and}~$P_{3}(m'_{3}<-1, m>0)$~\textrm{to}~$P_{1}(m'_{1}<-1, m>0)$\footnote{
      We define $m'_{i}\equiv m'\mid_{P_{i}}$.}\\

      The point $P_{1}$ is a stable one in the range $0<m<1/2$ provided that $m'_{1}<-1$, whilst $P_{3}$
      is always a saddle point. The curve $m(r)$ must intersect\footnote{Since the assumption $g'(R)>0$ leads to
      a monotonic function $r(R)$ and then, a single valued $m(r)$, hence, we do~not consider a multivalued
      $m(r)$.} the line $m=-r-1$ with a derivative $m'_{3}>-1$ or $m'_{3}<-1$ for leaving the matter epoch,
      and entering the accelerated epoch with $m'_{1}<-1$. Theoretically, the transition $P_{3}(m'_{3}>-1, m>0)$
      to $P_{1}(m'_{1}<-1, m>0)$ is possible as is shown in Figure~\ref{Figa} labeled Class~$I$ solutions. However, the
      transition from $P_{3}(m'_{3}<-1, m>0)$ to $P_{1}(m'_{1}<-1, m>0)$ is~not possible, these solutions are
      labeled as Class $VII_{a}$ in Figure~\ref{Figb}. Nevertheless, there is a special case, namely, when
      $P_{3}$ and $P_{1}$ are solutions of the model with the same root $r$, in which one has $m'_{1,3}<-1$.
      Hence, it demonstrates an acceptable cosmological solution, and we indicate this type of solution as Class~$II$ (Figure~\ref{Figa}).
\end{itemize}
\begin{itemize}\label{item2}
\item $P_{3}(m'_{3}>-1,m>0)$~\textrm{to}~$P_{2}(m'_{2}>-1)$~\textrm{in~Regions}~$
\mathcal{A},~\mathcal{D},~\mathcal{C}$~\textrm{and~to}~$P_{2}(m'_{2}<-1)$~\textrm{in~Region}~$\mathcal{B}$\\

    This class includes two classes of solutions. In the first class, there is no
    connection between $P^{(0)}_{3}$ and $P_{2}$ in the regions $\mathcal{A},
    \mathcal{D}$ and $\mathcal{C}$, which we call them as Class~$VII_{b}$ drawn in
    Figure~\ref{Figb}. All solutions with whether an improper transition (transition
    from unallowable regions) or without connection with the matter point $P^{(0)}_{3}$
    fall in this class. In the second class, it is possible to connect $P^{(0)}_{3}$
    with $m'_{3}>-1$ to $P_{2}$ with $m'_{2}<-1$ in the region $\mathcal{B}$ which we
    depict an example of these solutions in Figure~\ref{Figa} labeled as
    Class~$III$ solutions. Note that, these classes of solutions also appear in the $f(R)$ gravity.
\end{itemize}
\begin{itemize}\label{item3}
\item $P_{3}(m'_{3}<-1, m>0)~\textrm{to}~P_{2}(m'_{2}>-1)~\textrm{in~Regions}~\mathcal{
A},~\mathcal{D},~\mathcal{C}~\textrm{and~to}~P_{2}(m'_{2}<-1)~\textrm{in~Region}~\mathcal{B}$\\

    Since $P_{3}$ is a saddle point irrespective of the value of $m'$, it can be
    connected to the point $P_{2}$ in the regions $\mathcal{A},~\mathcal{D}$ and
    $\mathcal{C}$ in which the solutions are illustrated as Class~$IV$ in Figure~\ref{Figa}.
    On the other hand, because we have $m'_{2,3}<-1$ in the region $\mathcal{B}$,
    there is no possibility to connect $P^{(0)}_{3}$ to $P_{2}$ in this region, these
    solutions are classified as Class~$VII_{c}$ in Figure~\ref{Figb}.
\end{itemize}
\begin{itemize}\label{item4}
\item $P_{3}(m'_{3}>-1, m>0)~\textrm{and}~P_{3}(m'_{3}<-1, m>0)~\textrm{to}~P_{8}(0<m(r=-2)<1)$\\

    In this last class, there are two situations that can lead to a stable accelerated
    epoch. In the first one, after leaving the matter point, the trajectories go to
    the final attractor at the point $P_{8}$, and we refer to them as Class~$V$
    (Figure~\ref{Figa}). However, in the latter one, before reaching at the final
    attractor, there is a ``false'' accelerating era in which the curve $m(r)$ does
    meet the line $m=-r-1$ in an unallowable region of the point $P_{1}$, in which
    there is no stable accelerated expansion. These solutions are labeled as Class~$VI$
    (Figure~\ref{Figa}).
\end{itemize}
\section{Case Studies}\label{case studies}
In this section, without loss of generality, we examine the
discussed classification scheme by considering some well--defined
specific theories for simplicity and then, investigate the
possible cosmological solutions. First of all, by
``well--defined'', we mean the corresponding models whose $m(r)$
curves can be derived explicitly with respect to $r$. We do~not go
into the details unless there would be some new cosmological
solutions with respect to the $f(R)$ gravity, though to complete
the discussion, we may mention the other solutions wherever it is
necessary. As the related models are determined by the behavior of
their curves $m(r)$, our task is to find the cosmological
solutions by exploring the properties of these curves. The
following discussed theories are of form $f(R,T)=g(R)+\sqrt{-T}$,
where the functionality of $\sqrt{-T}$ is enforced by the
conservation law, equation (3.6). As mentioned in
Refs.~\cite{Sec.IV.1, Sec.IV.2}, some $f(R)$ gravity models cannot
pass the necessary criteria in order to have an acceptable
cosmological history, e.g., the lack of a deceleration expansion
period to admit a standard structure formation~\cite{Sec.IV.3}, or
a quick transition from radiation era to the late--time
acceleration or the lack of a connection between matter era and
the late--time acceleration era. In this respect, the authors of
Refs.~\cite{Sec.IV.1, Sec.IV.2,amend} have shown that theories of
the form $f(R)=\alpha R^{n}$ and $f(R)=R^{p}\exp{(qR)}$ do~not
lead to a connection between the standard matter era and
accelerated attractors. In addition to these difficulties, the
authors of Ref.~\cite{amend} have numerically shown that the
models of type $f(R)=R^{p}[\log{(\alpha R)}]^{q}$ and
$R^{p}\exp{(q/R)}$ suffer from a non--standard matter era for some
initial values. In the former one, the matter era is~not
effectively dominant and in the latter one, the standard matter
era is replaced by the $\phi$MDE epoch\rlap.\footnote{The
$\phi$--matter--dominated--epoch ($\phi$MDE) has been introduced
in Ref.~\cite{Sec.IV.4}.}\
 Hence, we reconsider the
following plausible models in the background of $f(R,T)$ gravity,
in order to find out whether these issues can be cured.
Incidentally, for $m$ as a constant parameter, definition
(\ref{parameterm}) gives $g(R)\propto R^{m+1}$, i.e. a power--law
function. Finally, at the end of this section, we briefly furnish
the comparison of the properties of solutions for the investigated
models in both $f(R,T)$ and $f(R)$ gravities in
Table~\ref{table:tableIII}.
\subsection{$f(R,T)=a R^{-\beta}+\sqrt{-T},~~~a>0,~~\beta\neq0$}
\begin{figure}[b]
\epsfig{figure=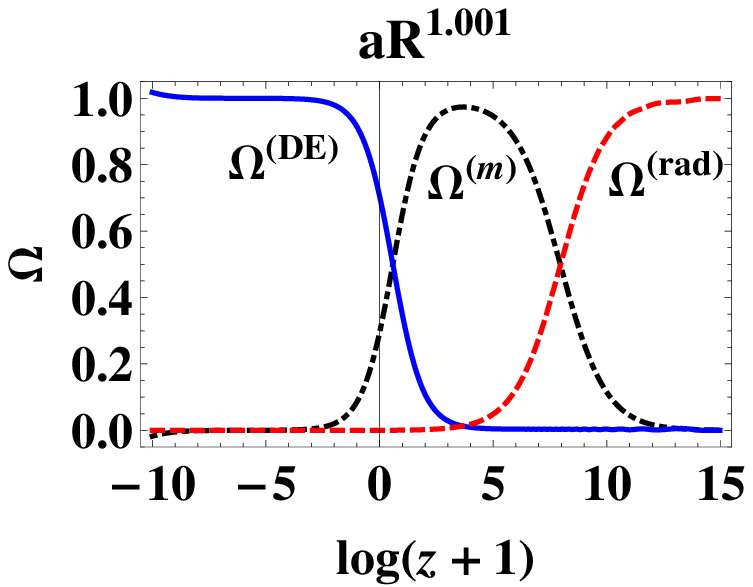,width=5.7cm}\hspace{3mm}
\epsfig{figure=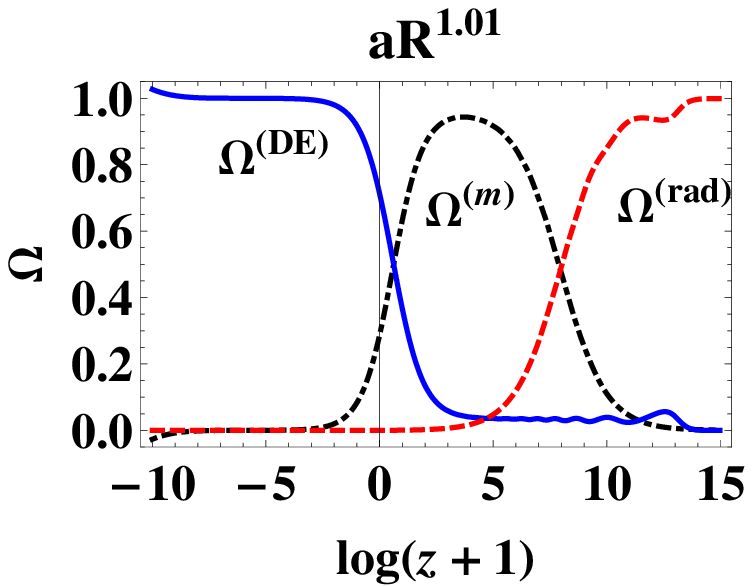,width=5.7cm}\hspace{3mm}
\epsfig{figure=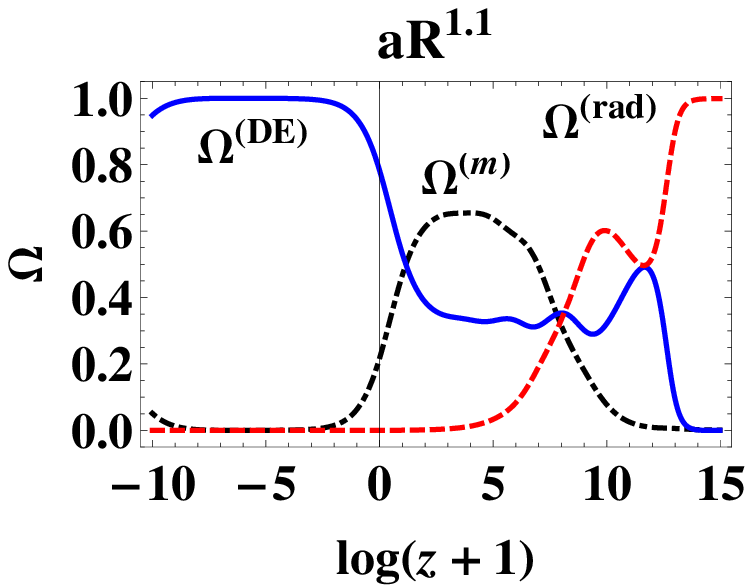,width=5.7cm}\vspace{4mm}
\epsfig{figure=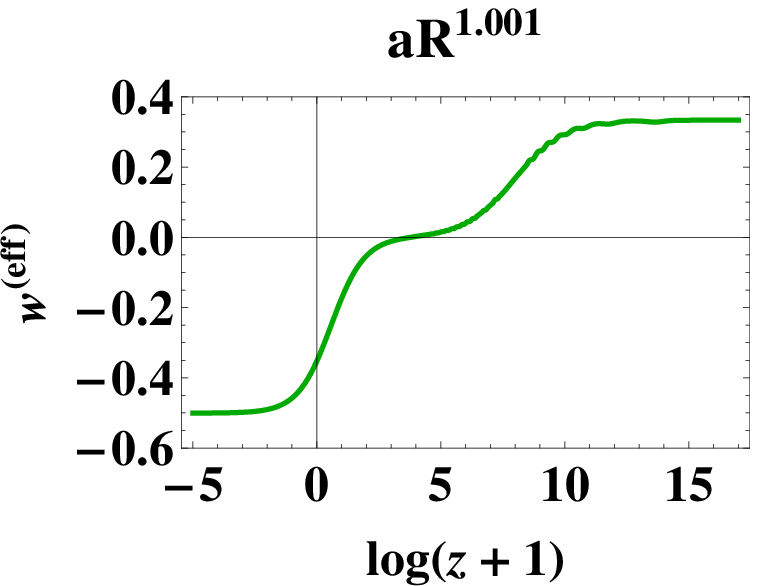,width=5.7cm}\hspace{3mm}
\epsfig{figure=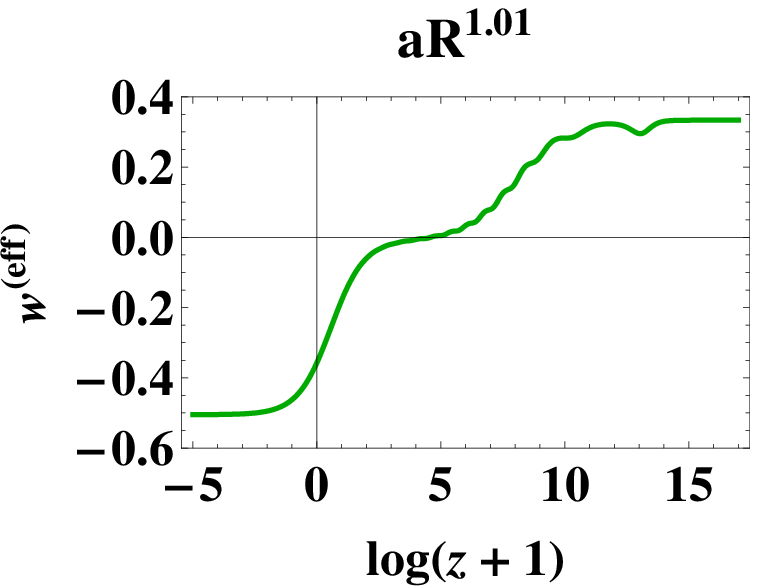,width=5.7cm}\hspace{3mm}
\epsfig{figure=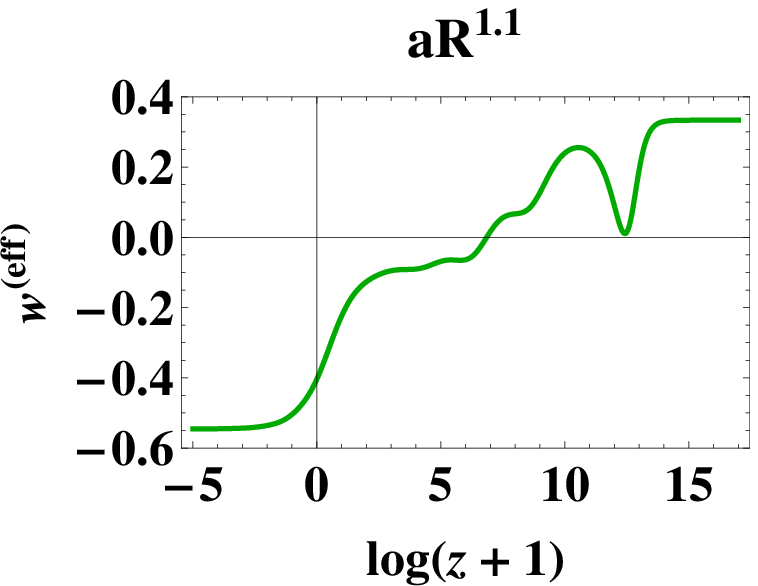,width=5.7cm}\vspace{4mm}
\epsfig{figure=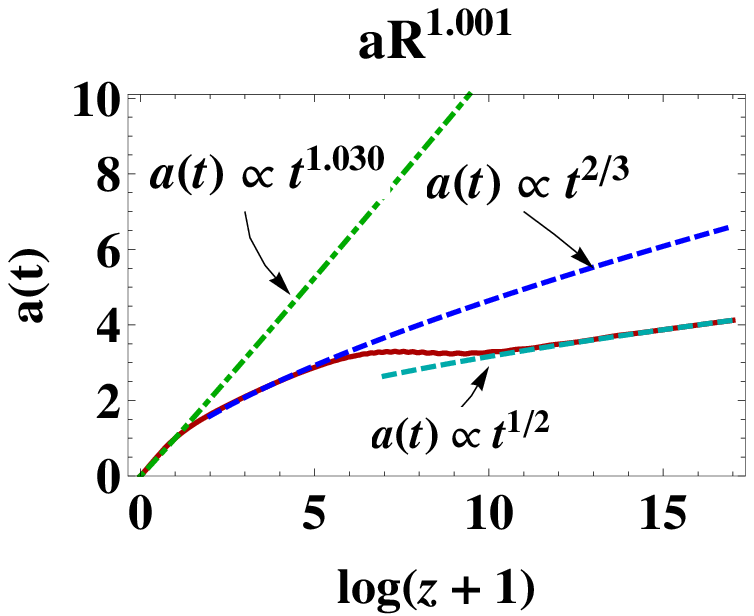,width=5.7cm}\hspace{3mm}
\epsfig{figure=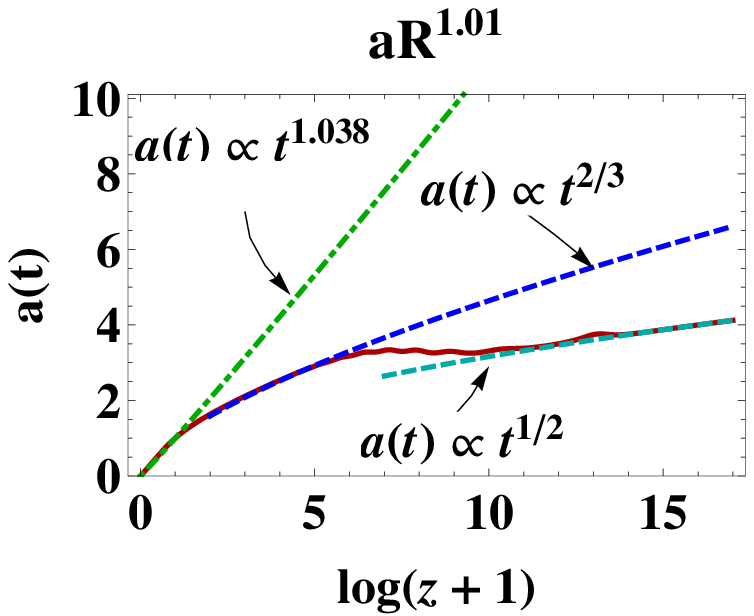,width=5.7cm}\hspace{3mm}
\epsfig{figure=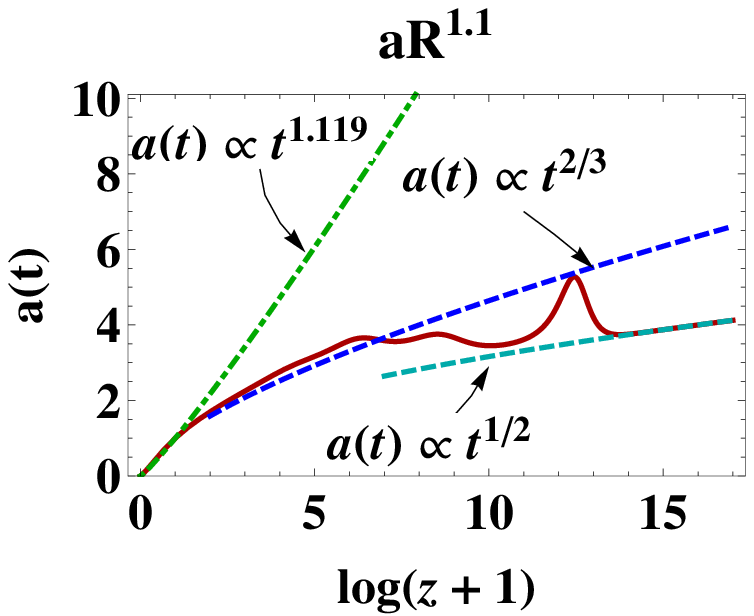,width=5.7cm} \caption{(color online).
\footnotesize {\textbf{Cosmological solutions of $f(R,T)=a
R^{-\beta}+\sqrt{-T}$ gravity.}} The numerical solutions with
$a>0$ for three values of $\beta$ are presented. The density
parameters for various ingredients are plotted in the first row,
the effective equation of state parameter in the second row and
the evolution of scale factor in the last row. The diagrams are
plotted for the initial values $x_{1}=10^{-4}$, $x_{2}=-10^{-4}$,
$x_{3}=-\beta \times 10^{-4}$, $ x_{4}=10^{-13}$ and $x_{5}=0.999$
corresponding to $z \thickapprox 2.42 \times 10^{7}$. The diagrams
are made to be consistent with $\Omega^{\textrm{(m)}}_{\textrm{0}}
\thickapprox 0.3$ and $\Omega^{\textrm{(rad)}}_{\textrm{0}}
\thickapprox 10^{-4}$ at the present epoch, however, they give
$-0.65<w^{\textrm{(eff)}}<-0.5$ instead of
$w^{\textrm{(eff)}}\rightarrow -1$. The peak of $\Omega^{(m)}$
decreases with the increment in $\beta$, i.e., as $\beta$
increases the diagrams get tangled up. Such disorderings are
indicated in the diagrams of $w^{\textrm{(eff)}}$ and in the
deviations of the behavior of the scale factor in the matter epoch
from its standard form $a\varpropto t^{2/3}$. The best solutions
are achieved for $\beta\rightarrow -1$.} \label{Figc}
\end{figure}

This theory gives $m(r)=-\beta-1$, which intersects the line
$m=-r-1$ at $r=\beta$. As $m(r)=0$ is valid only for $\beta=-1$,
hence, the condition ${\mathcal M}(r)=0$ must be satisfied for all
values of $\beta$ except for $\beta=-1$. In this case, because we
have $x_{3}=\beta x_{2}$, the system reduces to a system with
three degrees of freedom, in which the eigenvalues of the points
$P_{1}$ and $P_{3}$ are given by the first three values in
(\ref{eig1}) and (\ref{eig3}), respectively. To be more exact,
$P_{1}$ is accelerated in $-1.50 < \beta <-1.43$ and $-1.43 <
\beta < -1$, where in the first range, we have a stable
accelerated epoch, and the second one determines a spiral stable
accelerated epoch. On the other hand, for $-1.43 < \beta < -1$,
this theory has a saddle matter era with a damped oscillation when
$m\rightarrow0^{+}$, and for the same root, $P_{1}$ is a spiral
stable accelerated point which means the corresponding models
belong to Class $II$ for categorization purposes. Therefore, in
the background of $f(R,T)$ gravity, this theory has a cosmological
solution with a standard matter--acceleration epoch sequence
unlike the $f(R)$ gravity. We illustrate three examples of this
theory in Figure~\ref{Figc} with the same initial values except
$x_{3}$. In this case, $r$ has a constant value with respect to
the time. The diagrams show some disturbances originating from the
deviation of the magnitude of $\beta$ from one, i.e., as $|\beta|$
deviates from one, more disturbances occur. The reason is that the
increase of the magnitude of $\beta$ leads to the growth of the
deviation of $m$ from zero, and this causes the increment in the
error of matter and radiation solutions of the system of equations
(\ref{minimal 1})--(\ref{minimal 4}) in turn. The diagrams smooth
by decreasing in the deviation, showing an appropriate succession
of the radiation--matter--accelerated expansion eras. These
examples have the point $P_{1}$ as an attractor solution with
$-0.65<w^{\textrm{(eff)}}<-0.5$. The diagrams are made in order to
have the present values
$\Omega^{\textrm{(m)}}_{\textrm{0}}\approx0.3$ and
$\Omega^{\textrm{(rad)}}_{\textrm{0}}\approx10^{-4}$.
\subsection{$f(R,T)= R^{p}\exp(qR)+\sqrt{-T}$,~~~$q\neq0$}
\begin{figure}[h]
\epsfig{figure=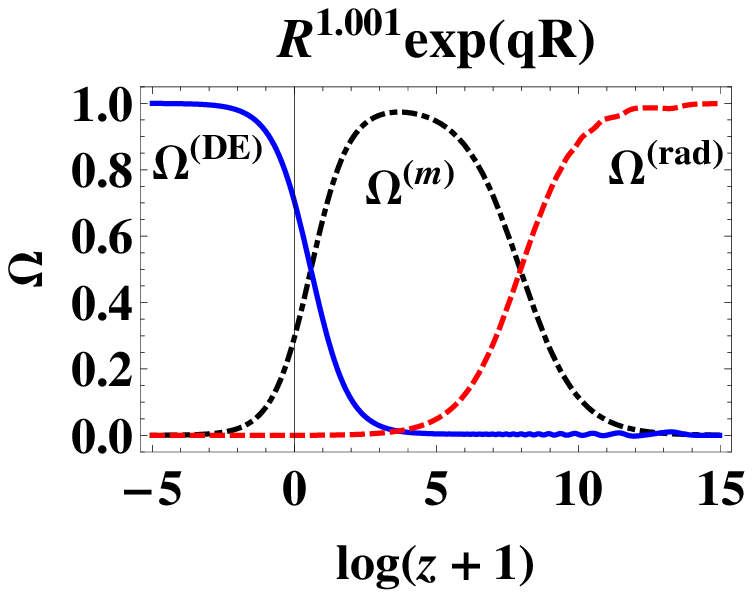,width=5.7cm}\hspace{3mm}
\epsfig{figure=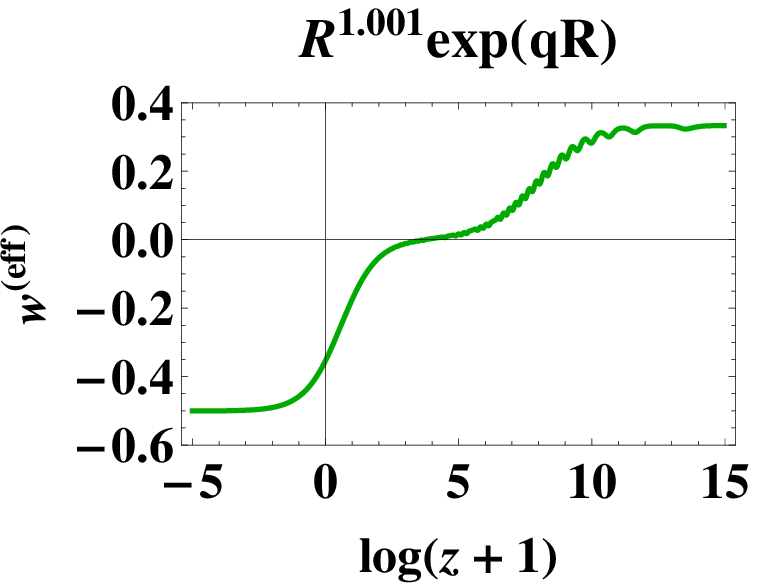,width=5.7cm}\hspace{3mm}
\epsfig{figure=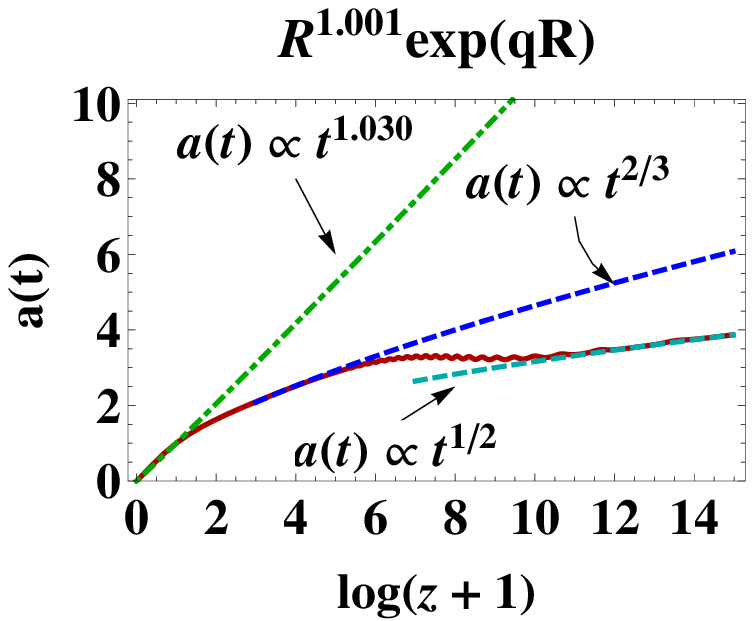,width=5.7cm}\ \caption{(color online).
\footnotesize {\textbf{Cosmological solutions of
$f(R,T)=R^{p}\exp(qR)+\sqrt{-T}$ gravity.}} The plots are
presented for $p=1.001$ and the initial values $x_{1}=10^{-4}$,
$x_{2}=-10^{-4}$, $x_{3}=1.002\times10^{-4}$, $ x_{4}=3.8 \times
10^{-13}$ and $x_{5}=0.999$ corresponding to $z \thickapprox 3.17
\times 10^{6}$. The diagram of $w^{\textrm{(eff)}}$ shows that the
final attractor solution is $P_{1}$. The present values of the
density parameters are extrapolated as
$\Omega^{\textrm{(m)}}_{\textrm{0}}\approx0.3$ and
$\Omega^{\textrm{(rad)}}_{\textrm{0}}\approx10^{-4}$.}
\label{Figd}
\end{figure}

In this theory, for $r\neq 0$, we get $m(r)=-r+p/r$, ${\mathcal
M}(r)=(p+r)/(p - r^2)$ and $m'(r)=-1-p/r^{2}$ which are
independent of $q$. For $r=-p$, the corresponding models do~not
satisfy the condition ${\mathcal M}(r)=0$ for $p=0$ and $p=1$,
however, for the other values of $p$, the two conditions hold. On
the other hand, for $p\approx0$, we have $m(r)\approx-r$, hence,
the condition for the existence of a matter solution,
$m(r\approx-1)\approx0^{+}$ is~not met and therefore, the pure
exponential models do~not have any cosmological solution.
Nevertheless, in addition to ${\mathcal M}(r)\approx 0$, in order
to have $m\rightarrow0^{+}$, the condition\footnote{More
precisely, $r\rightarrow -1^{-}$.} $r\approx -p$ for $p\rightarrow
1^{+}$ must hold. Actually, for this theory, we have $r=-1-qR$,
hence, we get $r\rightarrow-1$ from the left--hand side only when
$R\rightarrow0^{+}$ with $q>0$. Since $m'_{3}(p\rightarrow
1^{+})<-1$ and $m'_{1}(-3/2<r<-1)<-1$, it is impossible to connect
$P_{3}^{(0)}$ to $P_{1}$ for two different roots of
$m(r_{i})=-r_{i}-1$. One exception occurs when $P_{1}$ for the
same root is the attractor solution, thus in such a case, there is
a cosmological solution which belongs to Class~$II$. In
Figure~\ref{Figi}, in the $(r,m)$--plane, we draw a plot for the
$m(r)$ curve for this theory with $p=1.001$. Also, to illustrate
the idea, we numerically depict interesting cosmological
quantities predicted by this theory in Figure~\ref{Figd} for
$p=1.001$ and the initial value $r_{i}=-1.002$ in order to have
the present values $\Omega^{\textrm{(m)}}_{\textrm{0}}\approx0.3$
and $\Omega^{\textrm{(rad)}}_{\textrm{0}}\approx10^{-4}$.
\subsection{$f(R,T)= R+\alpha R^{-n}+\sqrt{-T}$,~~~$n\neq0$}\label{modelC}
\begin{figure}[h]
\epsfig{figure=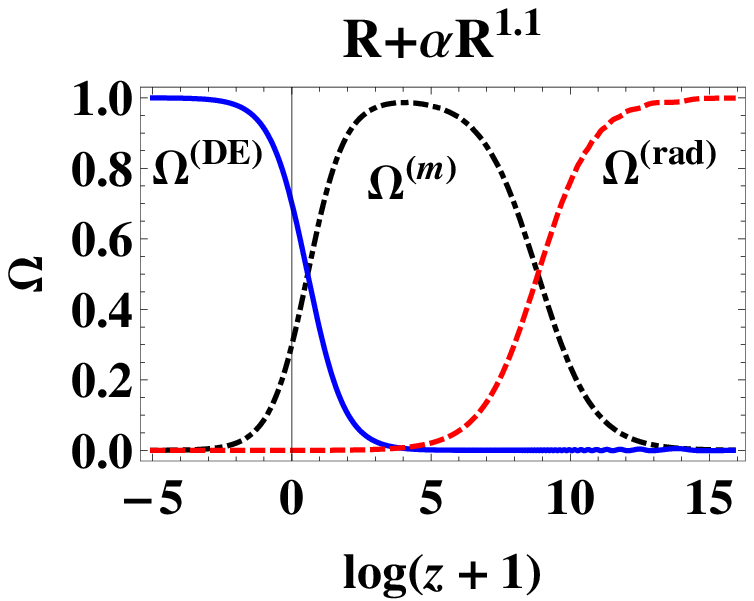,width=5.7cm}\hspace{3mm}
\epsfig{figure=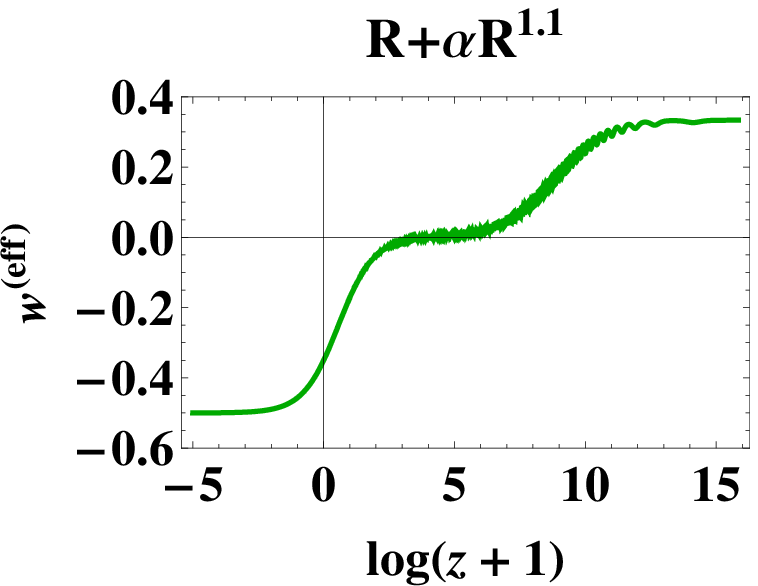,width=5.7cm}\hspace{3mm}
\epsfig{figure=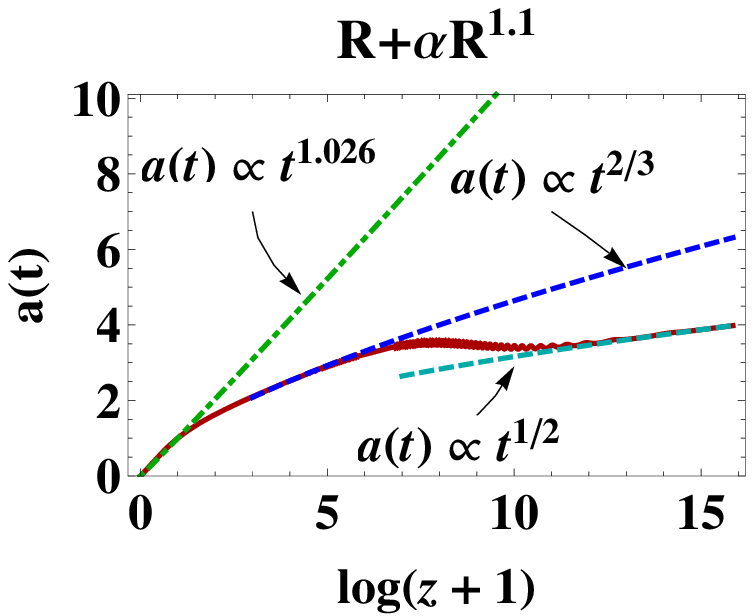,width=5.7cm}\vspace{8mm}
\epsfig{figure=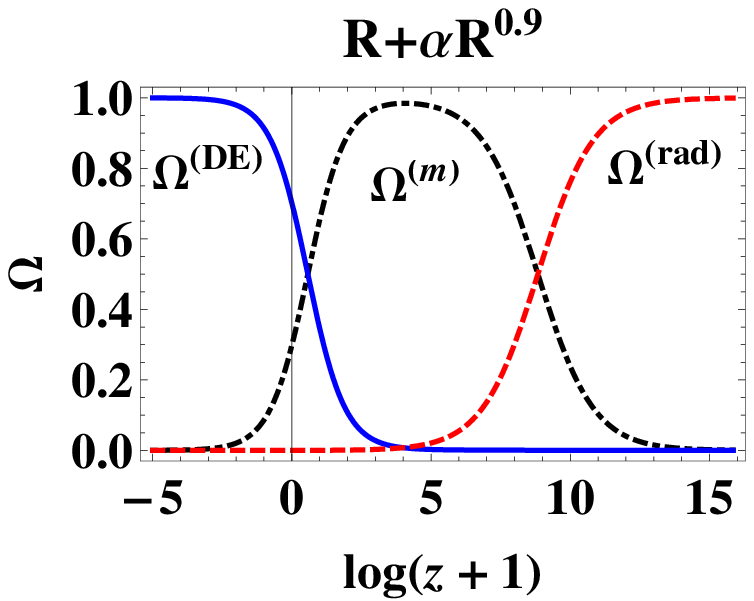,width=5.7cm}\hspace{3mm}
\epsfig{figure=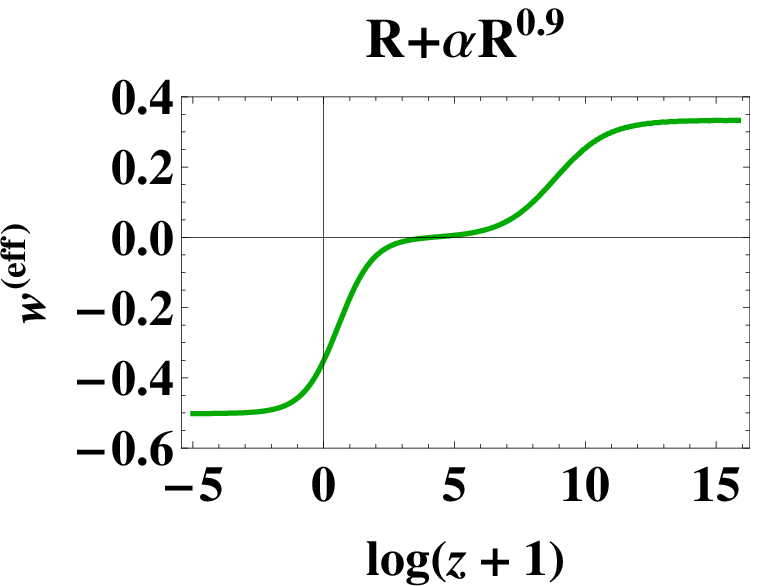,width=5.7cm}\hspace{3mm}
\epsfig{figure=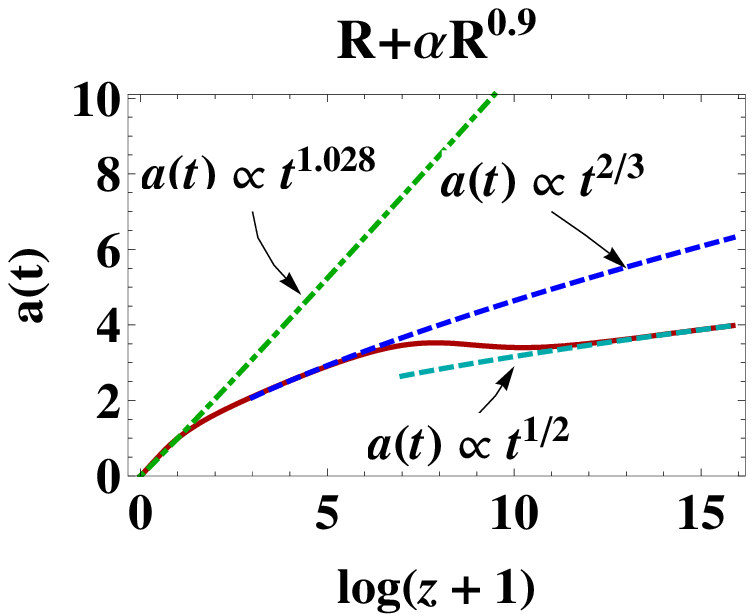,width=5.7cm}\vspace{8mm}
\epsfig{figure=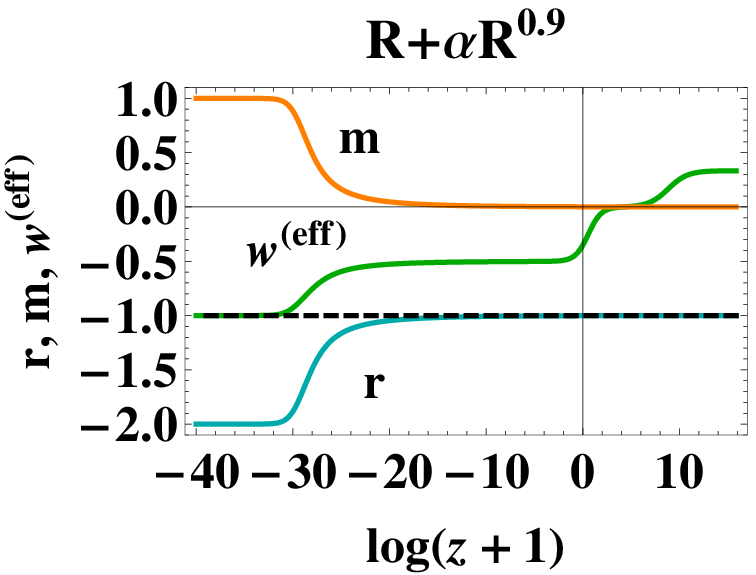,width=5.7cm}\ \caption{(color online).
\footnotesize \textbf{Cosmological solutions of $f(R,T)=R+\alpha
R^{-n}+\sqrt{-T}$ gravity.} The diagrams are obtained for $n=-1.1$
and the initial values $x_{1}=10^{-4}$, $x_{2}=-10^{-5}$,
$x_{3}=1.0008 \times 10^{-5}$, $ x_{4}=10^{-13}$ and $x_{5}=0.999$
corresponding to $z\thickapprox 7.65 \times 10^{6}$. Hence, the
model matches with the present observational data
$\Omega^{\textrm{(m)}}_{\textrm{0}}\approx0.3$ and
$\Omega^{\textrm{(rad)}}_{\textrm{0}}\approx10^{-4}$; however, its
$w^{\textrm{(eff)}}$ converges to value $-0.5$ instead of $-1$.
There are desirable successions of radiation--matter--acceleration
phases. The scale factor evolution curve has the asymptotic form
of $a\varpropto t^{1/2}$ at high--redshifts, and behaves as
$a\varpropto t^{2/3}$ when the matter becomes dominant. The model
with $n=-0.9$ is plotted for the same initial values except for
$x_{2}=-10^{-4}$ and $x_{3}=1.00002 \times 10^{-4}$. This model
shows a transition from a temporal acceleration epoch to the final
attractor in the vicinity of $P_{8}$, hence, it belongs to
Class~$VI$.} \label{Fige}
\end{figure}

For this theory, we obtain $m(r)=-n(1+r)/r$ and ${\mathcal
M}(r)=1-r/n$\,\rlap,\footnote{In obtaining the equation ${\mathcal
M}(r)=1-r/n$, we assume $r\neq-1$; however, after removing the
ambiguity at $r=-1$, it gives ${\mathcal M}(r=-1)=1+1/n$.}\ which
show that the condition ${\mathcal M}(r)=0$ is satisfied only for
$r=n$, that in turn gives $m\neq0$. On the other hand, the
corresponding models contain the matter point $P_{3}^{(0)}$ when
$r=-1$, which means that only models with $n=-1$ can be accepted.
However, we describe the properties of solutions for the values of
$n$ approaching $-1$ in the following models (where in these cases
we have ${\mathcal M}(r=-1)\approx0$).
\begin{itemize}
\item Models With $n\rightarrow-1^{-}$

In these cases, the equation $m(r_{i})=-r_{i}-1$ has two roots,
i.e., $r_{1,2}=-1,n$. Generally, we have $m'(r)=n/r^2$, as a
result, for the initial values $r_{i}<\sqrt{|n|}$, we have
$m'_{3}<-1$ and $m'_{1}>-1$, which denote $P_{1}$ to be a saddle
point. Thus, since $m(r=-2)=-n/2$, these solutions accept the
de~Sitter point $P_{8}$ as the final attractor after a transition
from the saddle point $P_{1}$. These models belong to Class~$VI$.
On the other hand, for $-\sqrt{|n|}<r_{i}<-1$, the point $P_{1}$
is a stable point. These solutions belong to Class~$II$. In
Figure~\ref{Fige}, we plot the related diagrams of $R+\alpha
R^{1.1}$ model. In this example, the initial value $r_{i}=-1.0008$
is applied in such a way that it chooses $P_{1}$ as the final
attractor and also gives the present values for the $\Omega$'s.
Also in Figure~\ref{Figi}, we present the $m(r)$ curve for this
example.

\item Models With $n\rightarrow-1^{+}$

For models with $n\rightarrow-1^{+}$, the initial conditions
$r_{i}>-\sqrt{|n|}$ are~not allowed, for, these conditions lead to
$m\approx 0^{-}$, which is physically ruled out. However, the
initial values $r_{i}<-1$ are allowed and give $m'_{1}>-1$.
Therefore, in these models, universe after passing a matter
dominated stage is trapped in a temporal accelerated expansion
state that is determined by $P_{1}$ and then, chooses $P_{8}$ as a
final de~Sitter attractor. These models belong to Class~$VI$. In
the numerical considerations, we have chosen $r_{i}=-1.00002$,
which results in an acceleration in a transient period by $P_{1}$,
then a permanent accelerated expansion by $P_{8}$; see
Figure~\ref{Fige}.
\end{itemize}

In the cases with $n\rightarrow-1^{-}$, cosmological solutions
exist only for $\alpha>0$ in the limit $R\rightarrow0$, whilst the
cases with $n\rightarrow-1^{+}$ have solutions provided that
$\alpha<0$ and $R\rightarrow\infty$ because $m=n(n+1)\alpha
R^{-n-1}/(1-n\alpha R^{-n-1})$. In the $f(R)$ gravity, these
models can have cosmological solutions only when $-1<n<0$,
however, in $f(R,T)$ gravity, in addition to these solutions,
there are acceptable solutions for $n\rightarrow-1^{-}$ as well.
\subsection{$f(R,T)= R^{p}[\log{(\alpha R})]^q+\sqrt{-T}$,~~~$q\neq0$,~~$\alpha>0$}
This theory has the following functions
\begin{align}
m(r)=\frac{(p+r)^2-qr(1+r)}{qr}~~~~~\mbox{\textrm{and}}~~~~~{\mathcal M}(r)=\frac{(p+r)^2}{(p+r)^2-qr(1+r)},
\end{align}
where $r\neq0$. The condition ${\mathcal M}(r)=0$ holds for
$r=-p$, however, only for $p=1$, we have $m(r)=0$ for
$r=-1$\rlap.\footnote{Note that, the existence of the matter point
for all types of this case is independent of $q$.}\ Incidentally,
for $p=1$ and $m(r=-2)=1-1/2q$, the point $P_{8}$ is a stable
accelerated attractor for $q>1/2$. Generally, in this theory,
there are three situations in which the matter solution
$m\rightarrow0^{+}$ can be obtained, namely,
\begin{align}
i)~\nonumber &q>\frac{1+r}{r},~~~r\rightarrow-1^{-} \Rightarrow ~~~m'>-1\\
ii)~\nonumber &q<0,~~~r\rightarrow-1^{-} \Rightarrow ~~~m'<-1\\
iii)~\nonumber &\frac{1+r}{r}<q<0,~~~r\rightarrow-1^{+} \Rightarrow ~~~m'>-1.\nonumber
\end{align}

The first situation shows that the corresponding models for
$0<q<1/2$ with $m'_{1,3}>-1$ lie in Class~$VII_{a}$. On the other
hand, $P_{2}$ cannot be the final attractor, for, the curve $m(r)$
does~not have any root in the regions $\mathcal{A}$,
$\mathcal{B}$, $\mathcal{C}$ and $\mathcal{D}$. However, for
$q>1/2$, the final attractor is $P_{8}$. The corresponding models
in the second situation, in which $m'_{1,3}<-1$, lie in Class~$II$
for the same root $r$ as represented in Figure~\ref{Fige}, but the
transition to $P_{2\mathcal{B}}$ does~not lead to a good
cosmological solution, which lies in Class~$VII_{c}$. In the last
situation, the range of $q$ gets narrowed as $r$ approaches to
$-1$, hence, it is of less importance to be studied.

\begin{figure}[h]
\epsfig{figure=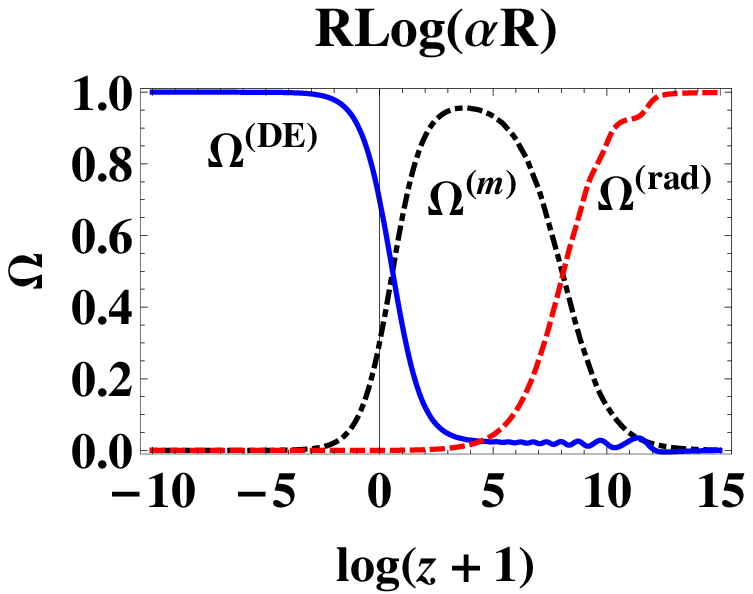,width=5.7cm}\hspace{3mm}
\epsfig{figure=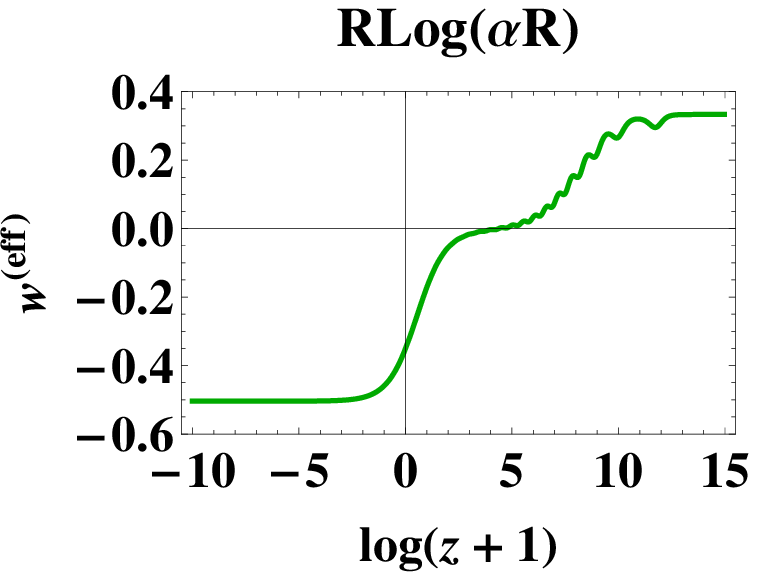,width=5.7cm}\hspace{3mm}
\epsfig{figure=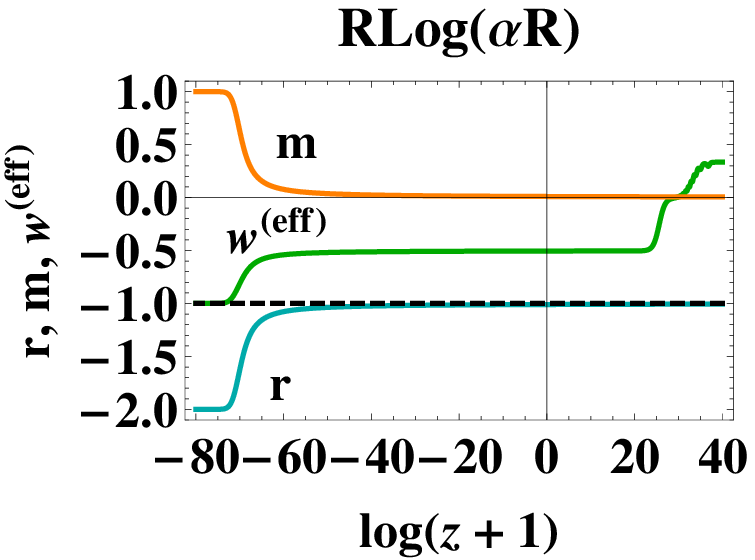,width=5.7cm}\vspace{4mm}
\epsfig{figure=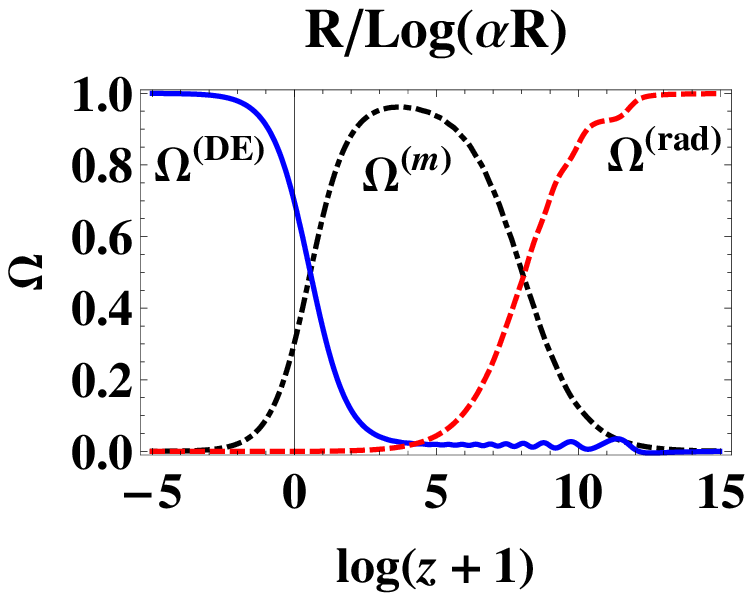,width=5.7cm}\hspace{3mm}
\epsfig{figure=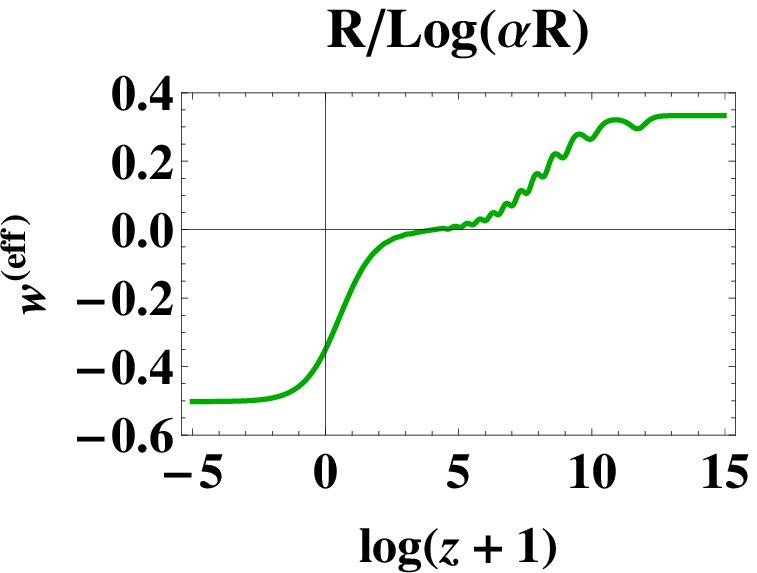,width=5.7cm}\hspace{3mm}
\epsfig{figure=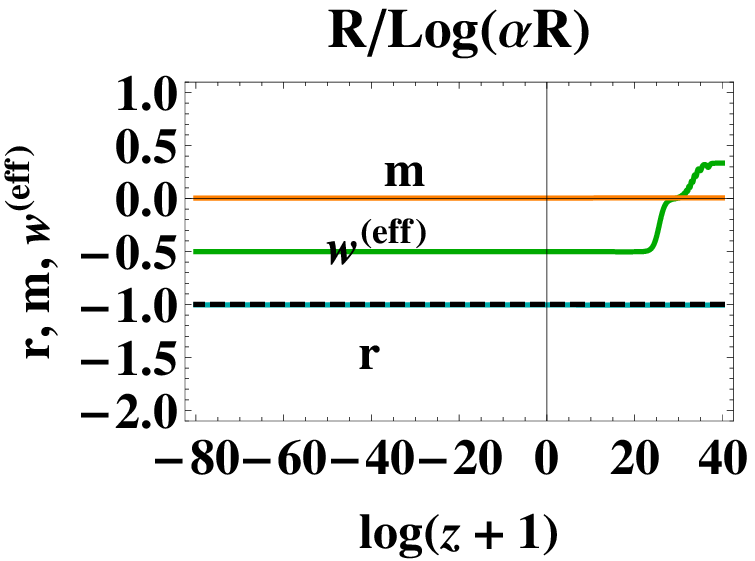,width=5.7cm} \caption{(color online).
\footnotesize \textbf{Cosmological solutions of
$f(R,T)=R(\log{\alpha R})^q+\sqrt{-T}$ gravity.} The density
parameter and the effective equation of state parameter diagrams
are plotted for two values $q=1$ and $q=-1$. The diagrams for $m$,
$r$ and $w^{\textrm{(eff)}}$ are drawn in a wide range of
redshift. The $\Omega$'s show admissible behaviors in both cases.
In the right panel, the first row diagram shows a transition from
a saddle accelerated epoch with
$w^{\textrm{(eff)}}\thickapprox-0.5$ to a stable one with
$w^{\textrm{(eff)}}\thickapprox-1$ for $q=1$. All three diagrams
support this result. Unlike $R \log{\alpha R}$, the theory $R /
\log{\alpha R}$ does~not show these transitions. We draw the
$m(r)$ curve for the model $R \log{\alpha R}$ in Figure~\ref{Figi}
which indicates that this theory belongs to Class~$VI$. The $m(r)$
curves of the theory first intersect the line $m=-r-1$ in an
unallowed region, then the line $r=-2$ as a final attractor. The
diagrams are plotted for the initial values $x_{1}=10^{-10}$,
$x_{2}=-10^{-7}$, $x_{3}=1.0058 \times 10^{-7}$, $ x_{4}=4 \times
10^{-13}$ and $x_{5}=0.999$ corresponding to $z\thickapprox 3.17
\times 10^{6}$ for both theories. The diagrams represent the
values of $\Omega^{\textrm{(m)}}_{\textrm{0}}\approx0.3$ and
$\Omega^{\textrm{(rad)}}_{\textrm{0}}\approx10^{-4}$ at the
present epoch.} \label{Figf}
\end{figure}

\begin{figure}[h]
\epsfig{figure=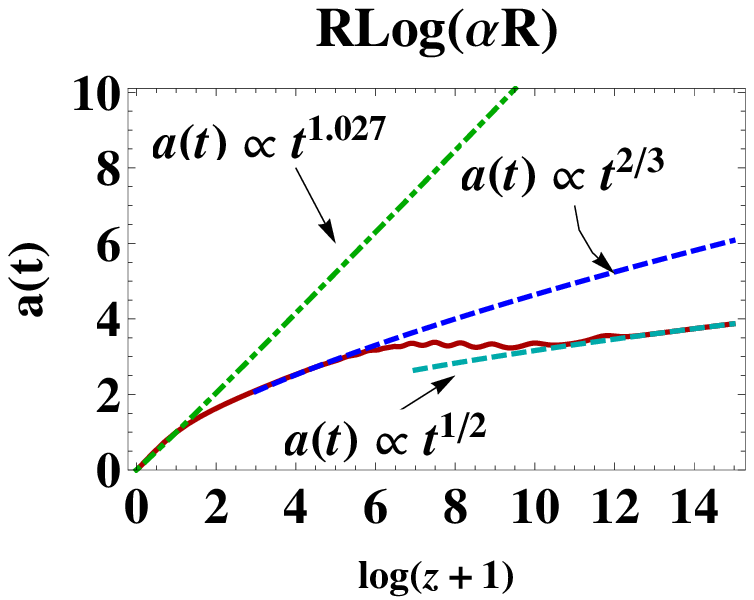,width=7cm}\hspace{4mm}
\epsfig{figure=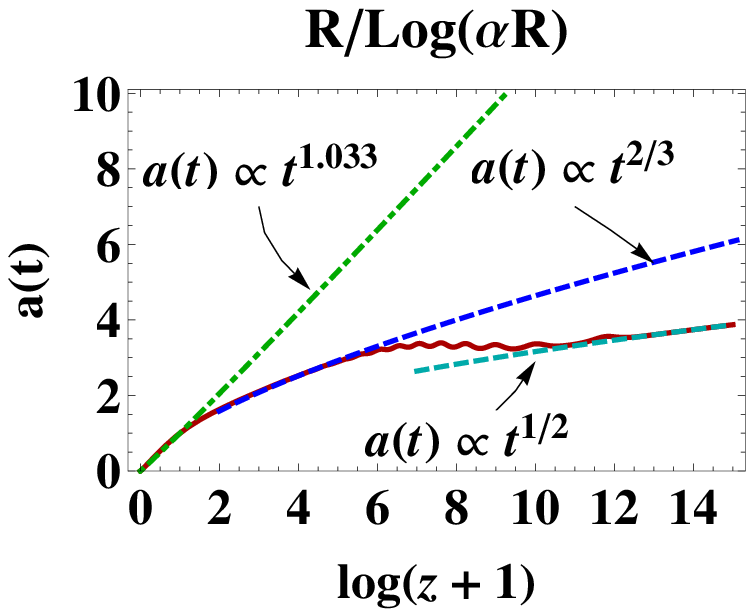,width=7cm}\hspace{4mm} \caption{(color
online). \footnotesize The scale factor evolution curves for the
two theories with $g(R)=R \log{\alpha R}$ and $g(R)=R /
\log{\alpha R}$ are depicted. The asymptotic lines show the
behaviors of the scale factor at the high--redshifts regime, in
the matter dominated epoch and at late times.} \label{Figg}
\end{figure}

In Figures~\ref{Figf} and \ref{Figg}, we plot two examples of such
cases for $q=\pm1$. The both examples show an acceptable
succession of the radiation--matter--accelerated expansion eras.
The theory $R\log{\alpha R}$ belongs to Class~$VI$ which has
$P_{8}$ as the final attractor. In addition to the curves of
density parameters for radiation, matter and acceleration eras,
the curves of $r\equiv-Rg'/g$, $m\equiv Rg''/g'$ and the effective
equation of state are depicted. These curves show a transition
from the saddle accelerated point $P_{1}$ to a stable de~Sitter
acceleration expansion phase after a long time. Figure \ref{Figf}
indicates that the curve $m(r)$ first intersects the line $m=-r-1$
in regions  $r\rightarrow-1^{-}$, in which $P_{1}$ is a saddle
point, then, intersects the line $r=-2$ where $P_{8}$ is a stable
point. Therefore, these solutions belong to Class~$VI$. Also,
$w^{\textrm{(eff)}}$ takes a transition from a non--phantom
accelerating era with the value
$w^{\textrm{(eff)}}\thickapprox-1/2$ to a de~Sitter epoch with
$w^{\textrm{(eff)}}\thickapprox-1$. Unlike this theory, the other
theory, i.e. $R/\log{(\alpha R)}$, has $P_{1}$ as the only
attractor as seen from Figure~\ref{Figf}. The latter theory
belongs to Class~$II$.

We conclude that the corresponding models of
$g(R)=R^{p}[\log{(\alpha R})]^q$ are cosmologically acceptable for
$p=1$ with $q<0$ and $q>1/2$ in the background of $f(R,T)$
gravity, whereas in the $f(R)$ gravity, the solutions exist only
in the range $q>0$.
\subsection{$f(R,T)= R^{p}\exp{(q/R)}+\sqrt{-T}$}
This theory has the relations
\begin{align}\label{mr5}
m(r)=-\frac{p+r(2+r)}{r}~~~~~\mbox{\textrm{and}}~~~~~{\mathcal M}(r)=\frac{p+r}{p+r(2+r)},
\end{align}
where $r\neq0$. The condition ${\mathcal M}(r)=0$ is satisfied
when $r=-p$ and for all values of $p$, except $p=0,1$. On the
other hand, the matter era only exists in $r=-p=-1$, hence, we
consider the theory in $r=-p$ when $p\rightarrow1^{+}$. In this
condition, we have $m'_{1}>-1$, therefore, the point $P_{1}$
cannot be stable whilst $P_{2}$ can be a stable accelerated point
in the region $\mathcal{C}$. Since $m''(r)=-2p/r^3$, thus the
point $(r\approx-1,m\approx0)$ is a minimum with a positive
concavity. Note that, for $r<-1$, we have $m(r)<-r-1$, which has
an asymptotic behavior in $r\rightarrow-\infty$ as $m(r)
\rightarrow -r$. As in this theory $r=(q/R)-1$, the latter
behavior occurs for $q<0$ and $R\rightarrow0^{+}$, which denotes
$P_{2\mathcal{C}}$ can be the final attractor. However, the
trajectories have already been trapped by $P_{8}$ in a finite $r$.
In this theory, like the corresponding models of $R\log(\alpha
R)$, before reaching the final attractor in $P_{8}$, there is a
short time interval in which the trajectories pass by $P_{1}$
(which is a saddle point). Thus, its corresponding models belong
to Class~$VI$.

\begin{figure}[h]
\epsfig{figure=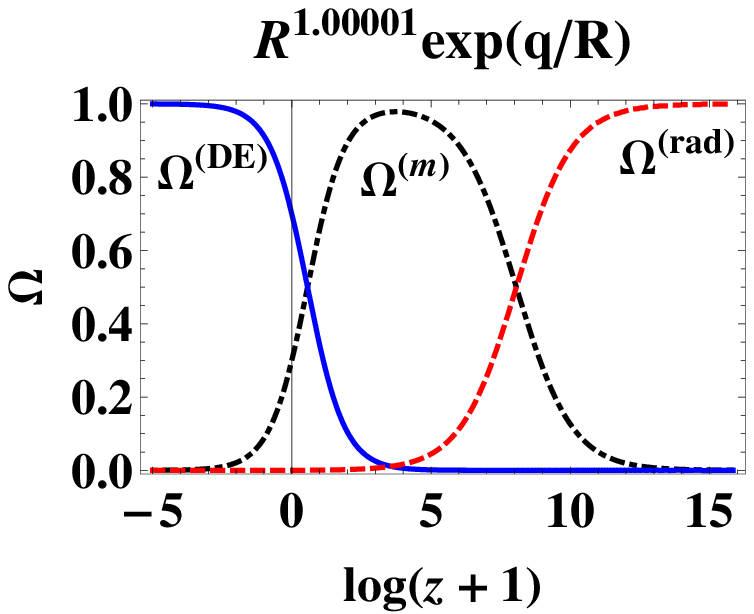,width=5.7cm}\hspace{2mm}
\epsfig{figure=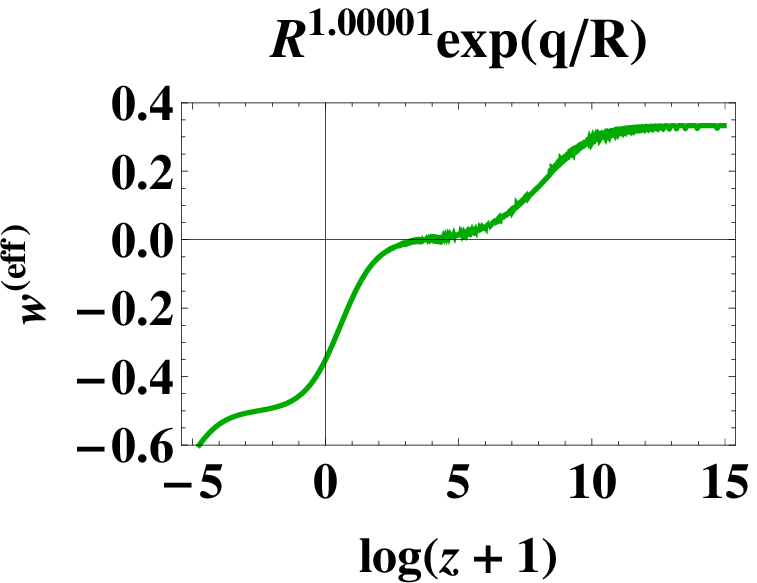,width=5.7cm}\hspace{2mm}
\epsfig{figure=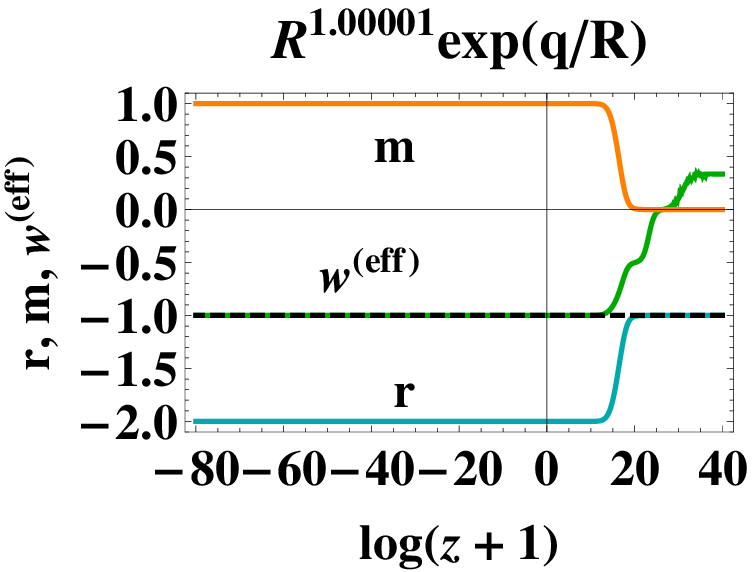,width=5.7cm}\vspace{4mm}
\epsfig{figure=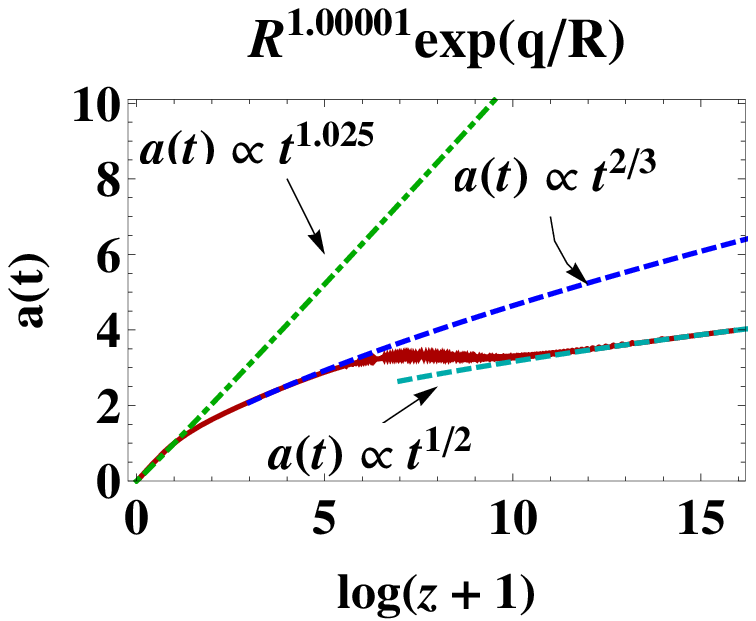,width=5.7cm} \caption{(color online).
\footnotesize \textbf{Cosmological solutions of
$f(R,T)=R^{p}\exp{(q/R)}+\sqrt{-T}$ gravity.} The plots are
provided for $p=1.00001$ and the initial values $x_{1}=10^{-5}$,
$x_{2}=-10^{-25}$, $x_{3}=1.00001 \times 10^{-25}$, $
x_{4}=10^{-15}$ and $x_{5}=0.9999$ corresponding to $z\thickapprox
3.53 \times 10^{7}$. In this case, there is always $m'_{1}>-1$,
which denotes $P_{8}$ as the final attractor.} \label{Figh}
\end{figure}

\begin{figure}[h]
\epsfig{figure=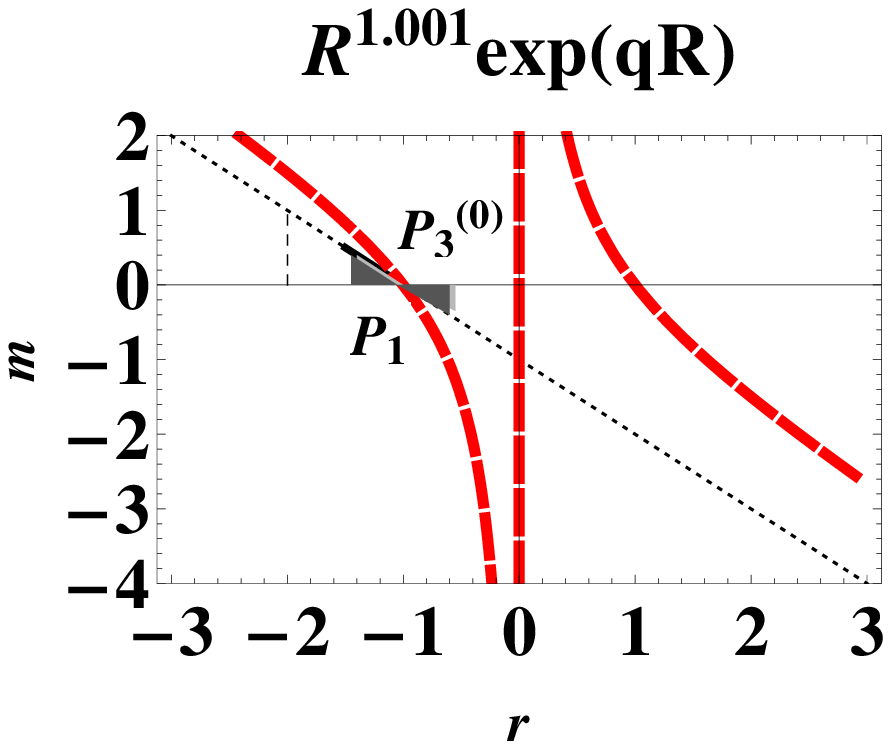,width=5.7cm}\hspace{3mm}
\epsfig{figure=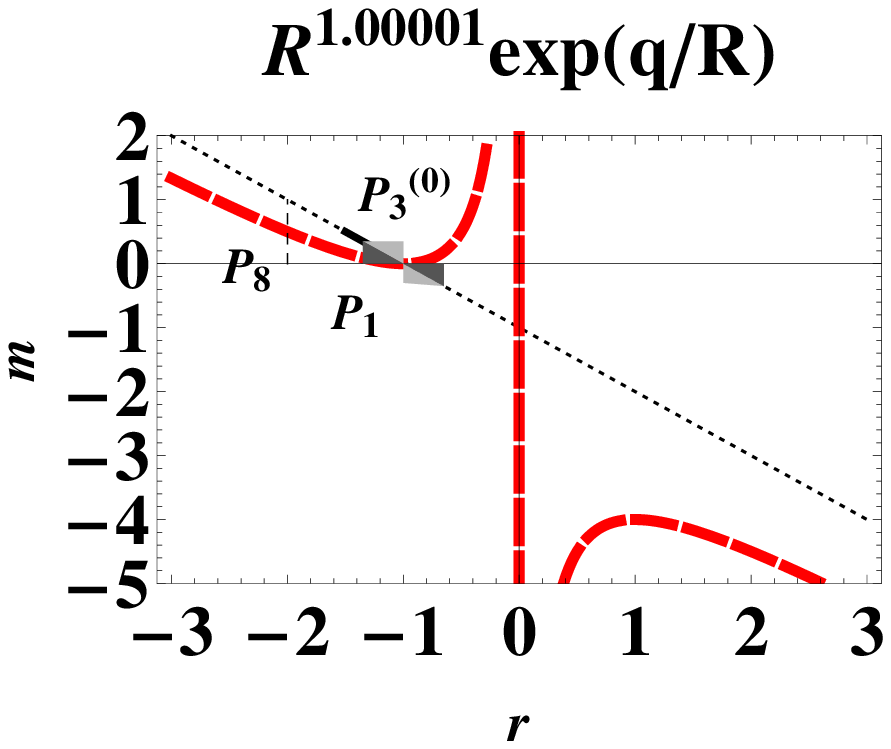,width=5.7cm}\hspace{3mm}
\epsfig{figure=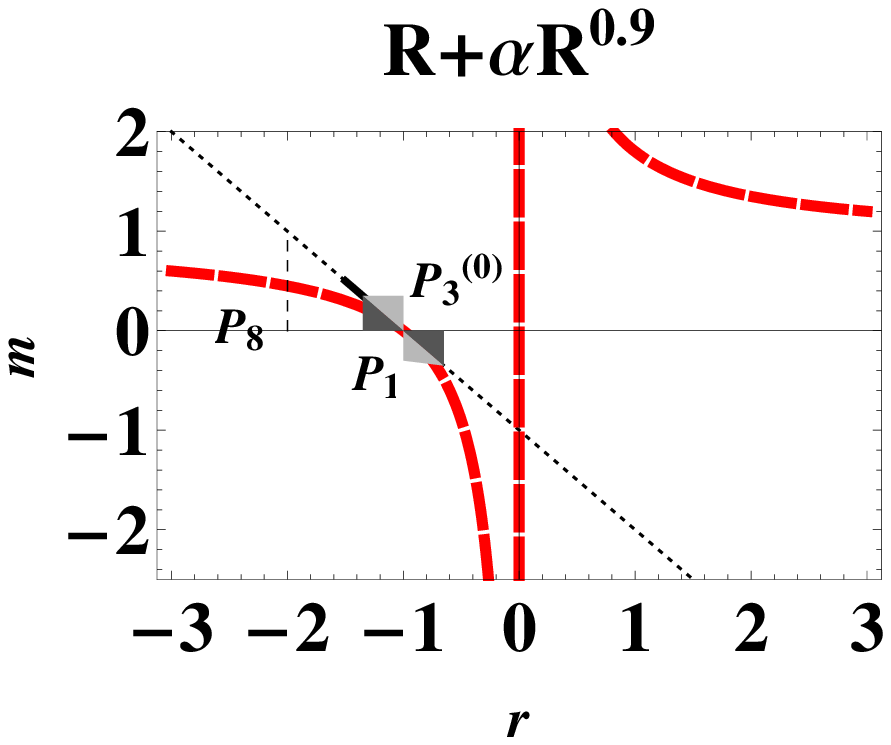,width=5.7cm}\vspace{4mm}
\epsfig{figure=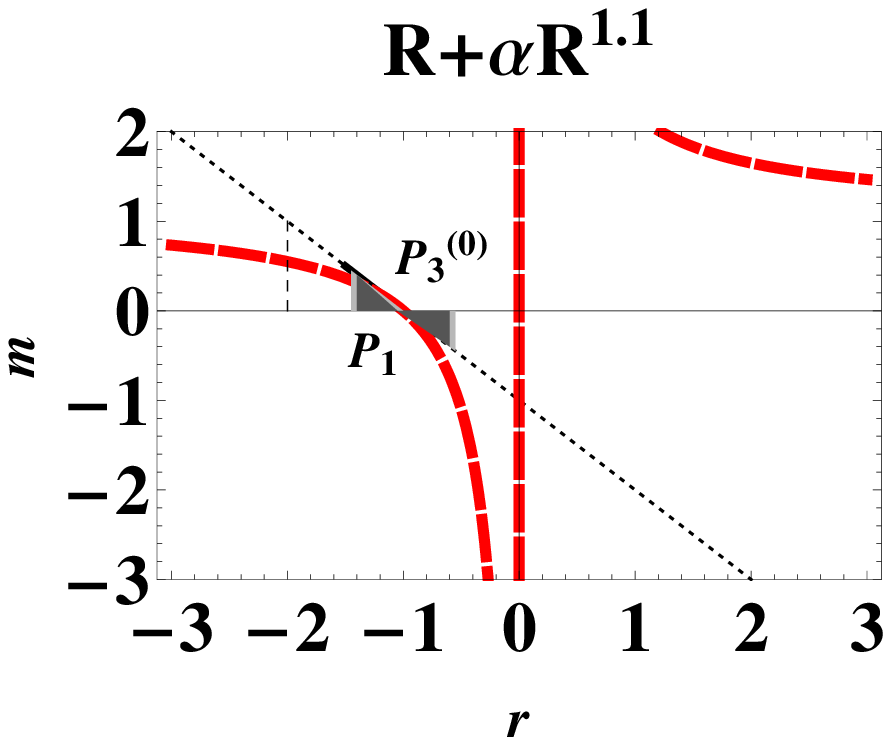,width=5.7cm}\hspace{3mm}
\epsfig{figure=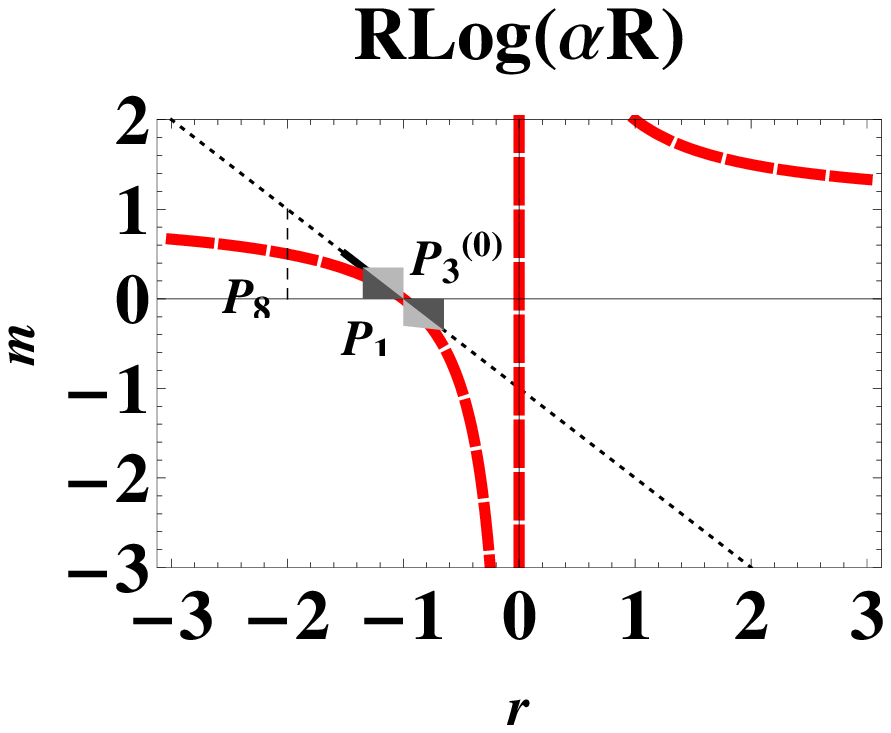,width=5.7cm}\hspace{3mm}
\epsfig{figure=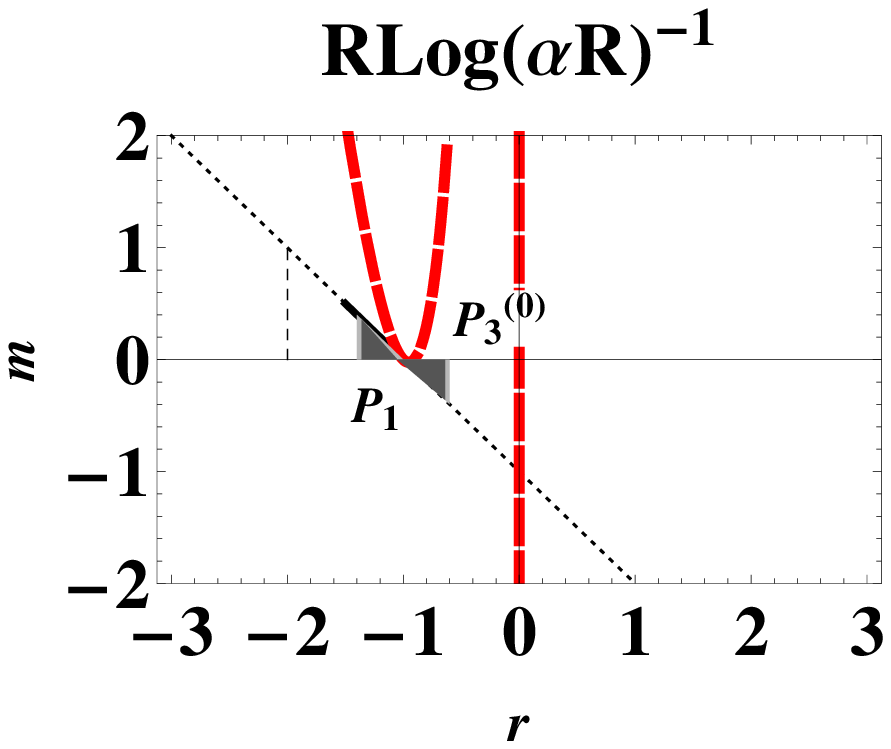,width=5.7cm}\ \caption{(color online).
\footnotesize \textbf{Theoretical curves for $m(r)$ for some
models in $f(R,T)$ gravity.} The $m(r)$ curves are illustrated for
some models corresponding to the represented classifications.
Contrary to the $f(R)$ gravity, the existence of the new point
$P_{1}$ and the new stability condition for $P_{3}$ ($P^{(0)}_{3}$
is a saddle point for both $m'<-1$ and $m'>-1$) bring about the
appearance of new acceptable solutions. The models for which
$P_{8}$ is the final attractor, and are indicated by the two
opposite triangles sets, belong to Class~$V$, and those for which
$P_{1}$ is the final attractor, and are indicated by the two
fitted triangles sets, belong to Class~$I$.} \label{Figi}
\end{figure}

As a result, the corresponding models, in general, have
cosmological solutions provided that $q<0$ for
$R\rightarrow\infty$. In Figure~\ref{Figh}, we depict a numerical
calculated example for these models provided we have the present
observed values for the density parameters. This example belongs
to Class~$VI$, as is obvious by comparing solutions of Class~$VI$
(Figure~\ref{Figa}) with the corresponding curve $m(r)$ in
Figure~\ref{Figi}.
\begin{table}[h]
\centering \caption{Cosmological solutions of $f(R,T)$ gravity
compared with the $f(R)$ gravity.}
\begin{threeparttable}
\begin{tabular}{|c||c|c|c|c||c|cc|}\hline\hline
Theory&\multicolumn{3}{c}{~~~~~~~~~~~$f(R,T)$ gravity} &&\multicolumn{2}{c}{$f(R)$ gravity}&\\[0.5ex]\hline
$g(R)$~Model                & Class $II$              & Class $VI$
   & Class $VII_{a}$ &Class $VII_{b}$   &Class $V$
   &Class $VII_{b}$&\\[0.5ex]\hline
$aR^{-\beta}$,~$a>0$        &$-1.43<\beta<-1$ &                  &
   &                  &
   & $-0.713<\beta<-1$&\\ [1ex]
$R^{p}\exp{(qR)}$             &$p\rightarrow 1^{+}$,~$q>0$       &
   &                 &$p\approx0$       &
   &$p=0,~p=1$&\\ [1ex]
$R+\alpha R^{-n}$           &$n\footnote{For models with
      $n\rightarrow-1^{-}$, we have $\alpha>0$ and for models with
      $n\rightarrow -1^{+}$, we have
      $\alpha<0$.}\rightarrow -1^{-}$&$n\rightarrow -1^{+}$,~$n\rightarrow-1^{-}$&&&$-1<n<0,~\alpha<0$&&\\[1ex]
$R^{p}(\log{\alpha R})^{q}$ &$p=1,~q<0$ &$p=1,~q>1/2$
   &$p=1,~0<q<1/2$    &
   &$p=1,~q>0$&$p\neq1$&\\[1ex]
$R^{p}\exp{(q/R)}$            &
   &$p\rightarrow 1^{+}$,~$q<0$ &                &
   &$p=1$   &$p\neq1$&\\[1ex]
\hline\hline
\end{tabular}
\end{threeparttable}
\label{table:tableIII}
\end{table}
\section{Pure Non--Minimal Case $f(R,T)=g(R)h(T)$}\label{non1}
Since we propose to investigate a pure non--minimal case in this
section, one should be more careful about the functionality of
$h(T)$. Indeed, in the vacuum state, we do~not want to have a null
Lagrangian. In a loose expression, there should be a solution for
the vacuum state (contrary with the strong version of the Mach
idea). Therefore, in this non--minimal case, the following
assumption is necessary, namely,
\begin{align}
\lim_{T \to 0} h(T)\neq0.
\end{align}
Furthermore, for simplicity, in this and the following sections,
we only consider the dust--like matter. Using the definitions
presented in Sec.~\ref{Fieldequation}, from (\ref{first}) and
(\ref{second}), we get
\begin{align}\label{nonminimal-1}
1+\frac{1}{6}\frac{g}{H^2 g'}-\frac{1}{6}\frac{R}{H^2}+\frac{\dot{g'}}{H g'}+\frac{\dot{h}}{Hh}=\frac{8
\pi G \rho^{(\textrm{m})}}{3 H^2 g' h}+\frac{gh' \rho^{(\textrm{m})}}{3H^2 g'h}
\end{align}
and
\begin{align}\label{nonminimal-2}
2\frac{\dot{H}}{H^2}+\frac{\ddot{g'}}{H^2 g'}+2\frac{\dot{g'}}{H g'}\frac{\dot{h}}{H h}
+\frac{\ddot{h}}{H^2 h}-\frac{\dot{g'}}{H g'}-\frac{\dot{h}}{H h}=-\frac{8 \pi G
\rho^{(\textrm{m})}}{H^2 g' h}-\frac{gh' \rho^{(\textrm{m})}}{H^2 g'h}.
\end{align}
Rewriting the first equation with respect to the defined variables
and parameters (\ref{varx1})--(\ref{varx3}) and
(\ref{parameters}), gives
\begin{align}\label{Nonminimal density}
\Omega^{(\textrm{m})}_{\rm p.n.}=1-x_{1}-x_{2}-x_{3}-s(3+2x_{2}),
\end{align}
where we have defined the corresponding pure non--minimal matter
density parameter as
\begin{align}\label{density definition}
\Omega^{(\textrm{m})}_{\rm p.n.}\equiv \frac{8 \pi G
\rho^{(\textrm{m})}}{3 H^2 g' h}.
\end{align}
In the above definition for $\Omega^{(\textrm{m})}_{\rm p.n.}$,
the existence of the function $h(T)$ warns us about the sign of
$\Omega^{(\textrm{m})}_{\rm p.n.}$. That is, as we have only
adopted that $g'(R)>0$, the density parameter
$\Omega^{(\textrm{m})}_{\rm p.n.}$ may obtain negative values due
to the appearance and functionality of $h(T)$. On the other hand,
if one physically demands that $\Omega^{(\textrm{m})}_{\rm p.n.}$
must be positive, then, the functionality of $h(T)$ will be
restrictive, e.g. an exponential function.

We have three dynamical equations for $x_{1}$, $x_{2}$ and
$x_{3}$; the equations for $x_{2}$ and $x_{3}$ are the same as
equation (\ref{minimal 2}) and (\ref{minimal 3}), respectively.
However, the equation for $x_{1}$ changes as
\begin{align}\label{nonminimal 1}
\frac{dx_{1}}{dN}=-1 + x_{1} (x_{1} - x_{3}) - 3 x_{2} - x_{3} + 3 s (2 x_{1} - x_{3} + 3 s).
\end{align}
In addition to the EOM constructed by equations (\ref{minimal 2}),
(\ref{minimal 3}) and (\ref{nonminimal 1}), we have a constraint
due to the energy--momentum conservation law. It is easy to check
that constraint (\ref{source}) leads to the following relations
among the parameters $n$ and $s$ with the other variables as
\begin{align}\label{nonmini. const. n}
n = \frac{x_{1} x_{3}}{3 m x_{2}}- \frac{1}{2}
\end{align}
and
\begin{align}\label{nonmini. const. s}
s = \frac{x_{1} x_{3}}{3 m x_{2}}+ \frac{1}{2},
\end{align}
which in turn lead to specify the functionality of $h(T)$ and also
to make a complicated relation between $g(R)$ and $h(T)$. Indeed,
further investigations indicate that $h(T)$ is a complicated
exponential function of $T$, which guaranties that
$\Omega^{(\textrm{m})}_{\rm p.n.}$ to be positive. Anyway, we have
a dynamical system with three variables and two constraints that
must hold. This system of equations accepts four fixed points
which are summarized in Table~\ref{puref}.
\begin{center}
\begin{table}[h]
\centering
\begin{threeparttable}
\caption{The fixed points solutions of $f(R,T)=g(R)h(T)$ gravity
without radiation.}
\begin{tabular}{l @{\hskip 0.1in} l@{\hskip 0.1in} l @{\hskip 0.1in}l @{\hskip 0.1in}l}\hline\hline

Fixed point     &Coordinates $(x_{1},x_{2},x_{3})$           &Parameter $s$  &$\Omega^{(m)}_{\rm p.n.}$      &$w^{\textrm{(eff)}}$\\[0.5 ex]
\hline
$P_{1}$&$\left(\frac{m(5-m-a_{m}\footnotemark[1])}{4(1+m)},~-\frac{11+17m+a_{m}}{8(1+m)^2},~\frac{11+17m+a_{m}}{8(1+m)}\right)$
  &$\frac{1}{12}\left(1+m+a_{m}\right)$&$0$&$-\frac{7+13m+a_{m}}{12(1+m)}$\\[0.75 ex]
$P_{2}$&$\left(\frac{m(5-m+a_{m})}{4(1+m)},~-\frac{11+17m-a_{m}}{8(1+m)^2},~\frac{11+17m-a_{m}}{8(1+m)}\right)$
  &$\frac{1}{12}\left(1+m-a_{m}\right)$&$0$&$-\frac{7+13m-a_{m}}{12(1+m)}$\\[0.75 ex]
$P_{3}$ &$\left(0,~-\frac{5}{4},~2\right)$&$0$&$0$&$-1$ \\[0.75 ex]
$P_{4}$ &$\left(-4,~-\frac{7}{4},~0\right)$&$0$&$0$&$\frac{1}{3}$ \\[0.75 ex]
\hline\hline
\end{tabular}
\label{puref}
\begin{tablenotes}
\small
\item[a] Where, $a_{m}\equiv\sqrt{-47 + 38 m + 121 m^2}$.
\end{tablenotes}
\end{threeparttable}
\end{table}
\end{center}

The most important point that can be observed is that the matter
density parameters of all the fixed points are zero. There is no
solution to describe a standard matter dominated era. Thus, we
do~not consider the general properties of their fixed points and
henceforth, their stabilities. However, the properties of each
point can be briefly summarized. The point $P_{1}$ is a
non--standard matter era, for, the condition
$w^{\textrm{(eff)}}=0$ is satisfied in $m=-2$, but we have
$\Omega^{(m)}_{\rm p.n.}=0$. $P_{1}$ is a de~Sitter point when
$m=0.6$. For this point, the non--phantom accelerating expansion
occurs when $0.48<m < 0.60$, and the phantom accelerating
expansion occurs when $m>0.6$. The point $P_{2}$ can expand
universe in the range $0.48<m<1.40$ in the non--phantom domain and
in the range $m < -1$ in the phantom domain. The point $P_{3}$ is
a special case of the point $P_{1}$ when $m=0.6$, which is a
de~Sitter point. And, contrary to our initial assumption for the
investigation of this theory, the point $P_{4}$ resembles a
radiation point, which is~not physically justified.
\section{Non--Minimal Case $f(R,T)=g(R)\left(1+h(T)\right)$}\label{non2}
Since a general Lagrangian $L=g_{1}(R)+g_{2}(R)h(T)$ makes the
calculations and the stability considerations more complicated, we
will just study the non--minimal case
$f(R,T)=g(R)\left(1+h(T)\right)$.

The following field equations are obtained, namely,
\begin{align}\label{mixed-1}
1+\frac{1}{6}\frac{g}{H^2 g'}-\frac{1}{6}\frac{R}{H^2}+\frac{\dot{g'}}{H g'}
+\frac{2\dot{h}}{H(1+2h)}=\frac{8 \pi G \rho^{(\textrm{m})}}{3 H^2 g' (1+2h)}+\frac{2gh' \rho^{(\textrm{m})}}{3H^2 g'(1+2h)}
\end{align}
and
\begin{align}\label{mixed-2}
2\frac{\dot{H}}{H^2}+\frac{\ddot{g'}}{H^2 g'}+\frac{\dot{g'}}{H g'}\frac{4\dot{h}}{H (1+2h)}
+\frac{2\ddot{h}}{H^2 (1+2h)}-\frac{\dot{g'}}{H g'}-\frac{2\dot{h}}{H (1+2h)}=-\frac{8 \pi G
\rho^{(\textrm{m})}}{H^2 g' (1+2h)}-\frac{2gh' \rho^{(\textrm{m})}}{H^2 g'(1+2h)}.
\end{align}
To get the dynamical equation from equations (\ref{mixed-1}) and
(\ref{mixed-2}), we need to define a new variable
\begin{align}\label{newvariable}
y\equiv \frac{h}{1+2h}.
\end{align}
Hence, the corresponding non--minimal matter density parameter
satisfies
\begin{align}\label{newvariablenew}
\Omega^{(\textrm{m})}_{n.}=1-x_{1}-x_{2}-x_{3}-2s(3+2x_{2})y,
\end{align}
where
\begin{align}\label{mixed density definition}
\Omega^{(\textrm{m})}_{n.}\equiv \frac{8 \pi G
\rho^{(\textrm{m})}}{3 H^2 g' (1+h)}.
\end{align}
Owing to these variables, the dynamical equations for $x_{1}$ and $x_{4}$ are derived as
\begin{align}\label{mixed nonminimal 1}
&\frac{dx_{1}}{dN}=-1 + x_{1} (x_{1} - x_{3}) - 3 x_{2} - x_{3} + 6 s (2 x_{1} - x_{3} + 3 s) x_{4},\\
&\frac{dx_{4}}{dN}=-3s x_{4}(1-2x_{4}),
\end{align}
where the corresponding equations for $x_{2}$ and $x_{3}$ remain
unchanged, i.e. equations (\ref{minimal 2}) and (\ref{minimal 3}).
In addition, the parameters $n$ and $s$ are also the same as in
equations (\ref{nonmini. const. n}) and (\ref{nonmini. const. s}),
which constrain the variables $x_{1}$, $x_{2}$ and $x_{3}$. The
fixed points of this system are represented in Table~\ref{mixedf}.
\begin{center}
\begin{table}[h]
\centering \caption{The fixed points of theory
                    $f(R,T)=g(R)\left(1+h(T)\right)$ without radiation.}
\begin{tabular}{l@{\hskip 0.1in}l@{\hskip 0.1in}l@{\hskip 0.1in}l}\hline\hline
Fixed point     &Coordinates $(x_{1},x_{2},x_{3},x_{4})$         &$\Omega^{(m)}_{n.}$      &$w^{\textrm{(eff)}}$\\[0.5 ex]
\hline
$P_{1}$&$\left(\frac{3m}{1+m},~-\frac{1+4m}{2(1+m)^2},~\frac{1+4m}{2(1+m)},~0\right)$&$\frac{2-m(3+8m)}{2(1+m)^2}$&$-\frac{m}{1+m}$\\[0.5 ex]
$P_{2}$&$\left(\frac{2(1- m)}{1 + 2m},~\frac{1-4 m}{m (1 + 2m)},~-\frac{(1-4m)(1+m)}{m (1 + 2m)},~0\right)$&$0$
  &$\frac{2-5m-6m^2}{3m(1+2m)}$\\[0.5 ex]$P_{3}$&$(-4,~5,~0,~0)$&$0$&$\frac{1}{3}$\\[0.5 ex]
$P_{4}$&$(0,~-1,~2,~0)$&$0$&$-1$\\[0.5 ex]
$P_{5}$&$\left(\frac{m(5-m-a_{m})}{4(1+m)},~-\frac{11+17m+a_{m}}{8(1+m)^2},~\frac{11+17m+a_{m}}{8(1+m)},~\frac{1}{2}\right)$
  &$0$&$-\frac{7+13m+a_{m}}{12(1+m)}$\\[0.5 ex]
$P_{6}$&$\left(\frac{m(5-m+a_{m})}{4(1+m)},~-\frac{11+17m-a_{m}}{8(1+m)^2},~\frac{11+17m-a_{m}}{8(1+m)},~\frac{1}{2}\right)$
  &$0$&$-\frac{7+13m-a_{m}}{12(1+m)}$\\[0.5 ex]
$P_{7}$&$(0,~-\frac{5}{4},~2,~\frac{1}{2})$&$0$&$-1$\\[0.5 ex]
$P_{8}$&$(-4,~\frac{7}{4},~0,~\frac{1}{2})$&$0$&$\frac{1}{3}$\\[0.5 ex]
\hline\hline
\end{tabular}
\label{mixedf}
\end{table}
\end{center}

In this theory, the point $P_{1}$ contains a matter dominated era,
and the other points give de~Sitter points and the accelerating
expansion domains. Nevertheless, it remains to ensure that the
matter point is a saddle point, and that we have a stable
accelerating point for the late time acceleration of universe. In
what follows, we only consider the properties of each fixed point
in turn to check these possibilities. More studies on the possible
cosmological solutions for some specific models can be carried
out; however, this is beyond the scope of this work.
\begin{itemize}
\item  The Point $P_{1}$

This point has the following eigenvalues
\begin{align}
-\frac{1}{2},~\frac{-3m+\sqrt{m(256 m^3+160 m^2 - 31 m-16)}}{4m(1+m)},~\frac{-3m-\sqrt{m(256 m^3+160 m^2 - 31 m-16)}}{4m(1+m)},~3(1+m').
\end{align}
The point $P_{1}$ is a stable point when $m'<-1$ and $0<m<0.346$,
and otherwise, a saddle point. If $m'=0$, it will be a saddle
point for all values of $m$. Also, when $m\rightarrow0^{+}$, it is
a saddle point provided that $m'>-1$. This point has a similar
property as the corresponding one in the $f(R)$ gravity.

\item  The Points $P_{2}$, $P_{5}$ and $P_{6}$

These three points can only play the role of attractor solutions
of the system, for, we have $\Omega^{(\textrm{m})}_{P_{2,5,6}}=0$.
The eigenvalues of $P_{2}$ are given as
\begin{align}\label{eigP2}
-4+\frac{1}{m},~~~\frac{-8m^{2}-3m+2}{m(1+2m)},~~~\frac{2 (1 - m^2)(1 +m')}{m(1 + 2 m)},~~~\frac{10m^2+3m-4}{6m(1+2m)}.
\end{align}
When $m<(-1-\sqrt{3})/2$ or $(-1+\sqrt{3})/2<m<1$, the point
$P_{2}$ can accelerate the expansion of universe in the
non--phantom domain, and when $-1/2<m<0$ or $m>1$, in the phantom
domain. There is no stable solution for the phantom accelerating
expansion, however, the stable non--phantom accelerating domains
are determined by
\begin{align}\label{mixed-p2-non-ph}
m'<-1,~~~~~~0.347\lesssim
m<1/2,~~~~~-1/3<w^{\textrm{(eff)}}<-0.260.
\end{align}

In the limit $m\rightarrow0$, the eigenvalues approximately read
as
\begin{align}\label{eigP2Lim}
\frac{1}{m},~~~\frac{2}{m},~~~\frac{2}{m}(1+m'),~~~-\frac{2}{3m},
\end{align}
which means that for both $m\rightarrow0^{+}$ and $m\rightarrow0^{-}$,
this point is a saddle one. Also, in the models with $m'=0$, the point $P_{2}$ is a saddle point for all values of $m$.

The point $P_{5}$ can accelerate the expansion of universe in the
non--phantom domain with $ -1< w^{\textrm{(eff)}} \lesssim-0.75$
for $0.486 \lesssim m < 0.6$, and in the phantom domain with
$w^{\textrm{(eff)}}<-1$ for $m>0.6$. $P_{5}$ is stable in the
first range provided that $m' < -1$, and when $m' > -1 $ in the
second range. Finally, $P_{6}$ is always a saddle point in the
non--phantom range $0.486 \lesssim m < 1.4$ and in the phantom
range $m < -1$ for all values of $m'$.

\item  The Points $P_{4}$ and $P_{7}$

The eigenvalues of $P_{4}$ are derived to be
\begin{align}\label{Mixedeigen8}
-3,~~~\frac{1}{2},~~~\frac{1}{2}\left(-3-\sqrt{25-\frac{16}{m}}~\right),~~~\frac{1}{2}\left(-3+\sqrt{25-\frac{16}{m}}~\right).
\end{align}
Clearly, $P_{4}$ is a saddle de~Sitter point. However, the
numerical calculations show that the point $P_{7}$ is a stable
de~Sitter solution for $0<m<1/2$.
\end{itemize}

We conclude this section with the assertion that the non--minimal
coupling Lagrangian $f(R,T)=g(R)\left(1+h(T)\right)$ can have
cosmological solutions in the form of transitions of $P_{1}$ to
any of the points $P_{2}$, $P_{5}$ or  $P_{7}$. Note that, the
fixed points $P_{3}$ and $P_{8}$ have $w^{\textrm{(eff)}}=1/3$
which means that they are~not physically justified in the absence
of radiation.
\section{concluding remarks}
In this work, we consider the cosmological solutions of $f(R,T)$
theory of gravity for a perfect fluid in a spatially flat,
homogeneous and isotropic background FLRW metric via the
$(r,m)$--plane analysis. We include the dust matter and radiation
in the action. We investigate some families of this theory that
can be written as a combination of a pure function of the trace,
e.g., $h(T)$, and a pure function of the Ricci scalar, e.g.,
$g(R)$, by the virtue of which one would be able to use $f(R,T)$
gravity as a modification of the $f(R)$ dark energy models. In
Ref.~\cite{amend}, by introducing two dimensionless parameters $r$
and $m$, the $(r,m)$--plane method has been employed to
parametrize the $f(R)$ function and simplify the calculations. In
this work, we extend their idea to the function $h(T)$ and
introduce another two new dimensionless parameters, namely $n$ and
$s$. With these definitions, we consider the cosmological
solutions of three general theories with the Lagrangians of
minimal, pure non--minimal and non--minimal couplings via the
dynamical systems approach. The conservation of the
energy--momentum tensor leads to a constraint equation that
relates $n$ to the other dynamical variables, and all acceptable
cosmological solutions must respect it.

In the minimal gravity, this constraint confines the function
$h(T)$ to a particular form, i.e. $h(T)=\sqrt{-T}+{\rm constant}$.
This theory gets specific values for the two new parameters, i.e.
$n=-1/2=-s$, and contains six classes of acceptable cosmological
solutions and three unacceptable ones with the following remarks,
particularly in comparison with the $f(R)$ gravity studied in
Ref.~\cite{amend}.
\begin{itemize}
\item In all of the solutions, the comparison of the value of the
slope of the $m(r)$ curve to $-1$ is of great importance. This
comparison determines the acceptability of the solutions from the
cosmological point of view, i.e., there should exist a succession
of a saddle radiation era, a saddle matter era and finally a
stable accelerated expansion era.

\item For all of the fixed points, one of the three conditions (\ref{setss}) must be satisfied.

\item There is a matter era solution, i.e.,
$P_{3}$, that is always a saddle point which exists for
$m\rightarrow 0^{+}$ with both $m'(r)<-1$ and $m'(r)>-1$. In the
$f(R)$ gravity, this fixed point is~not allowed for $m'(r)<-1$.

\item There is an important fixed point, i.e.
$P_{1}$, with the property $\Omega^{(m)}=0$ which acts as a stable
accelerated expansion point, in addition to the one that already
exists in the $f(R)$ gravity. This fixed point is the final
attractor in most models of the minimal coupling theory. However,
the relevant conditions for this point are
\begin{align}
    \nonumber\ m'(r)<-1~~~~~\mbox{and}~~~~~0<m<1/2.
\end{align}

\item There is a saddle point that indicates a ``false'' matter era
whose scale factor does~not behave like the one of matter era
(actually, its scale factor behaves as $t^{1/2}$ instead of
$t^{2/3}$). This point, which also appears in the $f(R)$ gravity,
can exist as the only matter point for some models.

\item There is a stable de~Sitter point that is the final
attractor of the theory. This point appears in the $f(R)$ gravity
too, and exists provided that
\begin{align}
    \nonumber\ 0<m<1~~~~\mbox{at}\quad r=-2.
\end{align}
\end{itemize}

The acceptable cosmological solutions must be a transition from a
saddle radiation era to a saddle matter era and finally be able to
be connected with an accelerated point as the final attractor,
provided that the matter domination should take long enough to
form cosmic structures. In principle, in this theory, we have two
matter points (one as ``standard'' and the other as
``non--standard'' points), two accelerated points and a de~Sitter
solution. Based on the existence of cosmological solutions, we
classify the acceptable solutions into six classes. Two of them
have the fixed point $P_{1}$ as the final attractor, two have
transitions to some regions of $P_{2}$, and for the last two,
$P_{8}$ acts as a de~Sitter solution. All these classes of
solutions are new ones with respect to $f(R,T)$ gravity, except
when the corresponding models have $P_{8}$ as a final attractor.
However, in $f(R,T)$ gravity, $P_{8}$ can be reached after passing
by $P_{1}$ for some periods. We briefly compare the properties of
solutions in terms of acceptable transitions for several specific
models in both $f(R,T)$ and the $f(R)$ gravities in
Table~\ref{table:tableIII}. Numerically, we show that theories
with $g(R)=a R^{-\beta}$, $g(R)=R^{p}\exp{(qR)}$, $g(R)=R+\alpha
R^{-n}$, $g(R)=R^{p} [\log{(\alpha R)}]^{q}$ and
$g(R)=R^{p}\exp{(q/R)}$ have proper sequences of the
radiation--matter--acceleration eras for some values of their
space parameters, which indicate that these theories deserve
further investigation. We show that for the corresponding models,
in which the cosmological trajectories advance to $P_{8}$, the
trajectories pass by $P_{1}$ before approaching to $P_{8}$. Also,
we have numerically checked that it is always possible to control
the duration in which the trajectories stay around $P_{1}$, and
the duration of the matter dominated era (the width of matter
density parameter in the related diagrams).

In $f(R,T)$ gravity with the minimal coupling, our investigated
models can present a standard cosmological history including
transient periods of radiation and matter domination followed by a
period of accelerated expansion domination, which can also give
the presently observed~\cite{Planck} contribution of the density
parameters $\Omega^{(m)}_0\simeq0.3$ and $\Omega^{\rm
(DE)}_{\textrm{0}}\simeq0.7$. Some of the models can explain the
accelerated expansion, via a dark energy with an effective
equation of state parameter of about $-1$. Though for some of the
other models, the trajectories are trapped in the point $P_{1}$and
hence, this effective parameter approaches to the value $-1/2$,
which contradicts the recent Planck results~\cite{Planck}. Also,
our models numerically suggest a power--law behavior of the scale
factors (near $z\simeq0$) of the form $a(t)\propto t^{n}$ for
$1.025<n<1.038$, which gives an accelerated epoch and leads to a
Hubble parameter of the form $H(z)\propto(1+z)^{1/n}$. These
results have been obtained numerically; however, a non--numeric
considerations can be performed to reconstruct these models with
constant parameters that are consistent with the present values of
$H_{0}$ and $\Omega^{(m)}_{0}$. Beyond these preliminary
considerations, one can also further constraint the models using
the type Ia supernovae measurements, the distance to the baryonic
acoustic oscillations and/or the position of the first peak in the
spectrum of anisotropies of CMBR observation. Indeed, one can
theoretically obtain the Hubble parameter $H(z)$ for each model
(which in addition to $H_{0}$ and $\Omega^{(m)}_{0}$, it may be a
function of the other constant parameters of the model) then,
performs the related calculations (e.g., the distance modulus of a
supernova at redshift $z$) using $H(z)$ and hence, compares
statistically the results with the available data (e.g., the
observed distance modulus of a supernova) to find out the best
values of the parameters of the model. Also, by a further step,
one can consider the model at the level of perturbation. That is,
by obtaining the effective gravitational constant (which, in
general, depends on the constant parameters of the model), one can
track the structure formation around the matter era and thus,
constrain the parameters of the model (e.g., the scalar
perturbations have been considered for some models in
Ref.~\cite{frtsp}). In a further study of $f(R,T)$ gravity, and in
an independent work, it would be our task to present the
observational constraints for our models.

The pure non--minimal theory has a few problems both in
fundamental and cosmological aspects. First of all, a Lagrangian
with the property $h(T)=0$ at $T=0$ does~not lead to the vacuum
solution, and actually one gets a null Lagrangian. In addition,
there is another problem in the cosmological regime, namely, there
is lack of a matter point. That is, all the fixed points have
$\Omega^{(m)}=0$; see Table~\ref{puref} for more details.

In the non--minimal theory, the corresponding fixed points consist
of the following.
\begin{description}
\item[(i)]
  The same
matter point as the minimal theory, but with different
eigenvalues. This point exists provided that $m\rightarrow 0^{+}$
and $m'(r)>-1$, which is the same as in the $f(R)$ gravity.

\item[(ii)]
 Three fixed points as stable
accelerated expansion solutions which are
\begin{align}
&\nonumber m'(r)<-1,~~~~~~0.347\lesssim m<1/2,~~~~~-1/3<w^{\textrm{(eff)}}<-0.260,\\
&\nonumber m'(r)<-1,~~~~~~0.486\lesssim m
<0.60,~~~~~-1<w^{\textrm{(eff)}}\lesssim-0.75,
\end{align}
in the non--phantom domain, and
\begin{align}
&\nonumber m'(r)>-1,~~~~~~ m >0.60,~~~~~w^{\textrm{(eff)}}<-1,
\end{align}
in the phantom domain.

\item[(iii)]
 A stable de~Sitter point which
exists provided that $0<m<1/2$.
\end{description}
Further considerations of the possible transitions and studies of
various models of this theory will be reported elsewhere.
\section*{Acknowledgments}
We thank the Research Office of Shahid Beheshti University G.C.
for financial support.

\end{document}